\documentclass[a4paper,fleqn,usenatbib]{mnras}

\usepackage{newtxtext,newtxmath}
\usepackage[T1]{fontenc}
\usepackage{color}
\usepackage{ae,aecompl}

\usepackage{graphicx}	% Including figure files
\usepackage{amsmath}	% Advanced maths commands
\graphicspath{{img/}}
\usepackage{adjustbox}

\newcommand{\teff}{$T_{\rm eff}$ }	% T_eff
\title[Metal-poor star kinematics]{Exploring the Galaxy's halo and very metal-weak thick disk with SkyMapper and Gaia DR2}

\author[G. Cordoni et al.]
{G. Cordoni,$^{1,2}$\thanks{E-mail: giacomo.cordoni@phd.unipd.it}
G. S. Da Costa,$^{2,3}$ D. Yong,$^{2,3}$ A. D. Mackey,$^{2,3}$ A. F. Marino,$^{1,4}$ 
\newauthor
S. Monty,$^{2}$ T. Nordlander,$^{2,3}$ 
J. E. Norris,$^{2}$ M. Asplund,$^{5}$ M. S. Bessell,$^{2}$
\newauthor
A. R. Casey,$^{6,7}$ A. Frebel,$^{8}$ K. Lind,$^{11}$ S. J. Murphy,$^{2,12}$
B. P. Schmidt,$^{2}$ X. D. Gao,$^{9}$ 
\newauthor
T. Xylakis-Dornbusch,$^{9,13} $A. M. Amarsi,$^{14}$ and A. P. Milone$^{1}$
\\
$^{1}$Dipartimento di Fisica e Astronomia ``Galileo Galilei'' --
  Universit\`{a} di Padova, Vicolo dell'Osservatorio 3, Padova, IT-35122\\
$^{2}$Research School of Astronomy and Astrophysics, Australian National University, Canberra, ACT 2611, Australia\\
$^{3}$ARC Centre of Excellence for All Sky Astrophysics in 3 Dimensions (ASTRO 3D), Australia\\
$^{4}$Centro di Ateneo di Studi e Attivita Spaziali ``Giuseppe Colombo'' -- CISAS, Via Venezia 15, I-35131 Padova, Italy\\
$^5$Max Planck Institute for Astrophysics,  Karl-Schwarzschild-Str. 1, D-85748 Garching, Germany \\
$^{6}$School of Physics and Astronomy, Monash University, Wellington Rd, Clayton, VIC 3800, Australia\\
$^{7}$Faculty of Information Technology, Monash University, Wellington Rd, Clayton, VIC 3800, Australia\\
$^{8}$Department of Physics and Kavli Institute for Astrophysics and Space Research, Massachusetts Institute of Technology, Cambridge, MA 02139, USA\\
$^{9}$Max-Planck-Institut f\"{u}r Astronomie, K\"{o}nigstuhl 17, D-69117, Heidelberg, Germany\\
$^{10}$Observational Astrophysics, Department of Physics and Astronomy, Uppsala University, Box 516, SE-75120 Uppsala, Sweden\\
$^{11}$Department of Astronomy, Stockholm University, AlbaNova, Roslagstullbacken 21, SE-10691 Stockholm, Sweden \\
$^{12}$School of Science, University of New South Wales, Canberra, ACT 2600, Australia\\
$^{13}$Landessternwarte, Heidelberg University, K\"{o}nigstuhl 12, D-69117, Heidelberg, Germany\\
$^{14}$Theoretical Astrophysics, Department of Physics and Astronomy, Uppsala University, Box 516 SE-75120 Uppsala, Sweden
}

\date{Accepted XXX. Received YYY; in original form ZZZ}

\pubyear{2020}

\begin{document}
\label{firstpage}
\pagerange{\pageref{firstpage}--\pageref{lastpage}}
\maketitle

\begin{abstract}
	% In the last few years the growing interest toward Extremely Metal-Poor stars ($\rm [Fe/H]<-2$) has culminated in a large number of spectroscopic and photometric surveys scanning the whole sky in the attempt of finding these stars.
	% These very rare stars are key ingredients to understanding the formation of the Milky Way, offering a unique perspective on the earliest phases of our infant Galaxy, and on the first generation of stars, the so-called \textit{Population-\rm III} stars. \\
%{\color{red} Version of 8 May, haven't check word count, 250 max.}\\
In this work we combine spectroscopic information from the \textit{SkyMapper survey for Extremely Metal-Poor stars} and astrometry from Gaia DR2 to investigate the kinematics of a sample of 475 stars with a metallicity range of $ -6.5 \leq \rm [Fe/H] \leq -2.05$ dex. 
Exploiting the action map, we identify 16 and 40 stars dynamically consistent with the \textit{Gaia Sausage} and \textit{Gaia Sequoia} accretion events, respectively. The most metal-poor of these candidates have metallicities of $\rm [Fe/H]=-3.31$ and $\rm [Fe/H]=-3.74$, respectively, helping to define the  low-metallicity tail of the  progenitors involved in the accretion events.  We also find, consistent with other studies, that $\sim$21\% of the sample have orbits that remain confined to within 3~kpc of the Galactic plane, i.e., |Z$_{max}$| $\leq$ 3~kpc. Of particular interest is a sub-sample ($\sim$11\% of the total) of low |Z$_{max}$| stars with low eccentricities and prograde motions.  The lowest metallicity of these stars has [Fe/H] = --4.30 and the sub-sample is best interpreted as the very low-metallicity tail of the metal-weak thick disk population.  The low |Z$_{max}$|, low eccentricity stars with retrograde orbits are likely accreted, while the low |Z$_{max}$|, high eccentricity pro- and retrograde stars are plausibly associated with the \textit{Gaia Sausage} system.  We find that a small fraction of our sample ($\sim$4\% of the total) is likely escaping from the Galaxy, and postulate that these stars have gained energy from gravitational interactions that occur when infalling dwarf galaxies are tidally disrupted.
\end{abstract}


%%%%%%%%%%%%%%%%%%%%%%%%%%%%%%%%%%%%%%%%%%%%%%%%%%

%%%%%%%%%%%%%%%%% BODY OF PAPER %%%%%%%%%%%%%%%%%%

\section{Introduction}\label{sec:intro}
	In the last decade, the astronomical community has experienced a renewal of interest in the properties of low-metallicity stars, particularly those with $\rm [Fe/H]<-2\,dex$\footnote{We will generally endeavour to follow the convention of \citet{beers2005} in that the terminology `very', `extremely', `ultra', etc, metal-poor indicates [Fe/H] $<$ --2.0, --3.0 and --4.0,  respectively.}.
	Motivated by the successful surveys of \citet{beers1992} and \citet{christlieb2008}, in recent years numerous spectroscopic \citep[e.g., SDSS, SEGUE, LAMOST, APOGEE;][]{york2000,yanny2009,cui2012,zhao2006,majewski2017} and photometric \citep[e.g., Pristine, SkyMapper;][]{starkenburg2017,wolf2018} surveys have been commissioned, scanning extensive sky-areas for these very rare and key objects. We refer to \citet[][their Section~1]{dacosta2019} for a more complete list of spectro-photometric surveys targeting low-metallicity stars.
	Not surprisingly, the underlying scientific motive is the understanding of the formation of our Galaxy, as well as other galaxies in the Universe.
	
	Specifically, the lowest metallicity stars observable at the present-day formed from gas enriched with the nucleosynthetic products from first generation  metal-free stars, the so-called Population-{\rm III} stars.  Studies of abundances and abundance ratios in  ultra- and extremely metal-poor stars can then yield constraints on the properties of the Pop~III stars, such as their masses, and on star formation processes at the earliest times \citep[e.g.,][]{frebel2015}.  Moreover, the kinematics of these stars can also provide much information on the events that occurred during the formation of the Milky Way (MW), which is believed to include both star formation in-situ and the accretion of lower-mass galaxies.  Indeed, together the abundances and kinematics of the lowest metallicity stars offer a distinct perspective on the earliest stages of the formation and evolution of the Milky Way, and by implication, of galaxies in general.%i.e., the first objects formed out of metal-free pristine gas right after the Big Bang, are thought to be massive and short-lived \citep[]{bromm2013, hirano2014}, so that they ``immediately'' enriched the environment with the products of their nucleosynthesis. The ejecta from these stars, which depend upon their properties, particularly their mass, are then incorporated into the subsequent generation(s) of low-metallicity stars. Those metal-poor stars therefore offer a unique perspective into the early stages of the formation and evolution of the Milky Way Galaxy (hereafter MW), (see, for example, the review by \citet{frebel2015}). 
	
	In terms of the formation of the MW, the most common scenario predicts that the most metal-poor stars will be found mainly in the Galactic halo and Bulge, as these components likely formed in the earliest stages of the MW's evolution
	\citep[e.g.][]{white2000,brook2007,tumlinson2010,elbadry2018}. In such a scenario relatively few, if any, very metal-poor stars are expected to lie in the MW disk as it formed at a later epoch after the settling into the plane of gas enriched by multiple generations of star formation  \citep[e.g.][]{hawthorn2016}.  However, recent kinematic results from surveys for the most metal-poor stars have cast doubt on this scenario, altering our understanding of the formation of the Milky Way. 
	For example, the recent studies of \citet{sestito2019a,sestito2020}, \citet{dimatteo2019} and \citet{venn2020} have revealed a new scenario where $\sim\,20\,\%$ of very metal-poor stars have orbits that are confined to within 3\,kpc of the MW plane; evidently the majority of these stars are not Galactic halo objects despite their low metallicities.
	
	In particular, \citet{sestito2019a} compiled a catalogue of 42 ultra metal-poor ([Fe/H] $\leq$ --4.0) stars from the literature and analyzed their orbital properties making use of Gaia DR2 proper motions
	\citep{gaiacollaboration2018}. They found that 11 out of 42 stars have prograde orbits that are confined to within 3~kpc of the Milky Way disk. Moreover, two of these MW-planar stars are found to be on nearly circular prograde orbits, and one is the star with the lowest overall metal content currently known \citep{caffau2011}.
	In the same fashion, \citet{dimatteo2019} investigated the kinematics of a sample of coincidentally the same number of low-metallicity stars drawn from the ESO Large Program ``First stars -- First nucleosynthesis''
	\citep{cayrel2004}. Their analysis also finds that $\sim 20\%$ of the stars show disk-like kinematics. They went on to consider a larger sample of stars covering a wider metallicity range and found consistent results.  \citet{dimatteo2019} then postulated the existence of an ``ultra-metal poor thick disk'' that is an extension to low metallicities of the Galaxy's thick disk population.
	
	\citet{sestito2020} carried out a similar kinematic analysis on a substantially larger sample, consisting of 1027 very metal-poor stars with [Fe/H] $\leq$ --2.5 selected from the 
	Pristine \citep{starkenburg2017,aguado2019} and LAMOST \citep{cui2012,li2018} surveys.  Again they find that almost 1/3rd of the stars in the sample have orbits that do not deviate significantly from the disk plane of the Galaxy.  They suggest that this implies that a significant fraction of the MW's metal-poor stars formed with the Milky Way (thick) disk. Moreover, they note that as a consequence, the history of the disk must have been sufficiently quiescent that (presumably old) metal-poor stars were able to retain their disk-like orbits to the present-day
	\citep{sestito2020}.
	
	\citet{venn2020} have also investigated the kinematics of metal-poor stars using a sample of 115 objects chosen from the Pristine survey \citep{starkenburg2017} that have been observed at high dispersion.  They find 16, out of 70, metal-poor stars whose orbits are confined to the vicinity of the Galactic plane, together with small numbers of stars that may have unbound orbits. They also identify stars whose orbital characteristics/actions are consistent with an origin in the \textit{Gaia Enceladus} \citep{helmi2018} accretion event.
	
	These somewhat unexpected results support the idea that the metallicity distribution of the Galaxy's thick disk does indeed possess a low metallicity tail, as first advocated by
	\citet{norris1985} and \citet{morrison1990}. Moreover, the proposed low metallicity tail would extend to lower metallicities than those authors suggested \citep[see also][]{chiba2000,beers2014}.
 
	The origin(s) of these metal-poor thick disk stars is, however, still uncertain, though the implications of their existence for the formation and evolution of the MW, and disk galaxies in general, are likely significant.  A number of different possibilities have been discussed \citep[e.g.][]{sestito2019a,sestito2020,dimatteo2019} including that the stars were accreted from small satellites once the MW disk had already formed, or that they represent low metallicity stars formed in the gas-rich building-blocks that came together to form the main body of the Galaxy's disk \citep[see also the theoretical simulations presented in][]{sestito2020b}.
	
	In this work we conduct a similar study to those mentioned above by exploiting the metallicity determinations from the \textit{SkyMapper Survey for extremely metal-poor stars} 
	\citep[see][]{dacosta2019}, together with Gaia DR2 astrometry
	\citep{gaiacollaboration2018},
	to investigate the dynamics of 475 very metal-poor ([Fe/H] $<$ --2) stars in the southern sky.  The wide extension in metallicity space, together with the relatively large number of stars, gives us a detailed view of the kinematic properties of these objects.  We also consider the potential connection of any of the stars in our sample with the MW accretion events,
	such as those designated \textit{Gaia Enceladus, Gaia Sausage} and \textit{Gaia Sequoia} that have been recently discovered in large scale analyses of Gaia DR2 data  \citep[e.g.][]{helmi2018,belokurov2018,myeong2019,mackereth2019}.  
	Such a connection has also been pursued in \citet{monty2019}.
	
	The paper is organized as follows: in Sections~\ref{sec:data} and \ref{sec:kin} we present the data set and the orbit determination procedure, respectively, while in \S~\ref{sec:results} and \S~\ref{sec:discussion} we present and discuss our results.  Specifically, in \S~\ref{sub:escapers} we discuss the small number of stars in our sample that appear not to be bound to the Galaxy.  The final section (\S~\ref{sec:conc})
	summarizes our findings.

\section{Data}\label{sec:data}
	The data set used in this work consists of 475 stars with metallicities ranging from $\rm[Fe/H]=-2.08$ to $\rm[Fe/H] < -6.5$ dex.
	It is composed as follows: 
	\begin{itemize}
		\item 114 giant stars with $-6.2 \leq \rm [Fe/H]_{1D,\,LTE} \leq -2.25$ dex. Of these stars 113 come from Yong et al. (in preparation), while the remaining star is the most-iron poor star for which iron has been detected: SMSS~J160540.18--144323.1 with $\rm [Fe/H]_{1D,\,LTE} = -6.2 \pm 0.2$ \citep{nordlander2019}. These stars originate with the extremely metal-poor (EMP) candidates discussed in \citet{dacosta2019} and all have been observed at high resolution, principally with the MIKE spectrograph \citep{bernstein03} at the 6.5m Magellan (Clay) telescope. We shall refer to these stars as the \texttt{HiRes} data set.
		
		\item 45 stars observed with the \texttt{FEROS} high-resolution spectrograph \citep{kaufer99} at the MPG/ESO 2.2-metre telescope at La Silla.  Again, these stars originated from the \citet{dacosta2019} sample. We removed from the analysis all the stars with $\rm [Fe/H]>-2$, and the stars in common with \texttt{HiRes} data set. The final count of stars belonging to this sub-sample is 38 and we label it as the \texttt{FEROS} data set.

		\item 122 stars from \citet{jacobson2015} which have $-3.97\leq {\rm[Fe/H]}\leq-1.31$ dex. These stars originated in the SkyMapper commissioning-era survey \citep[see][]{dacosta2019}, and were also observed at high-dispersion with the MIKE spectrograph at Magellan. As for the \texttt{FEROS} sample, we removed 7 stars with $\rm [Fe/H]_{1D,\,LTE} > -2$ and the single star in common with the \texttt{HiRes} data set. However, as discussed in \S \ref{sub:orb}, there appears to be an issue with the radial velocities for the stars observed by \citet{jacobson2015} during one specific Magellan/MIKE run, namely 2013 May 28 -- June 01.  As a result, we have removed the stars observed in that run that lack a radial velocity from Gaia DR2 and which had not been already discarded. The final sub-sample used here is then composed of 91 stars and we refer to it as the \texttt{Jacobson+15} sub-sample. 

		\item 17 stars from \citet{marino2019} with metallicity $-3.26\rm<[Fe/H]_{1D,\,LTE}<-1.71$ dex. The spectra of these stars were obtained with the Keck HIRES high-resolution spectrograph \citep{vogt94}. After the removal of 2 stars present in the \texttt{Jacobson+15} sub-sample, and 2 stars with $\rm [Fe/H]_{1D,\,LTE} > -2$, we retain 13 stars.
		This sub-sample is referred to as the \texttt{Marino+19} data set.

		\item 362 giant star candidates from \citet{dacosta2019} with either $\rm [Fe/H]_{\rm fitter}$\footnote{$\rm [Fe/H]_{\rm fitter}$ is determined from the low-resolution spectra as described in \citet{dacosta2019}.}$<-3.0$, or $-3.0 \leq \rm [Fe/H]_{\rm fitter} \leq -2.5$ and $g_{\rm SkyMapper}<13.7$ mag. The radial velocities from the low-resolution spectra lack sufficient precision for our analysis, so the list of stars was cross-matched with Gaia DR2 to obtain radial velocities. A total of 195 stars were retained after the cross-match.  These stars are referred to as the \texttt{LowRes} data set.
		
		\item 24 Ultra Metal-Poor giant stars $({\rm [Fe/H]}\leq-4)$ from \citet{sestito2019a}, included to increase the number of UMP stars in the full sample and to provide a consistency check on our procedures. We have specifically selected only known giants from their sample for consistency with the SkyMapper derived samples, which are giant dominated.
		We refer to \citet{sestito2019a} for a detailed description of the data set but we note it includes the star SMSS J031300.36-670839.3, which has 
		${\rm [Fe/H]_{3D, NLTE}}<-6.5$ \citep{keller2014,bessell2015,nordlander2017}.  This data set is referred to as the \texttt{Sestito+19} sub-sample.
	\end{itemize}

\begin{figure}
		\centering
		\includegraphics[width=8cm, trim={2cm 0cm 2cm 0cm}, clip]{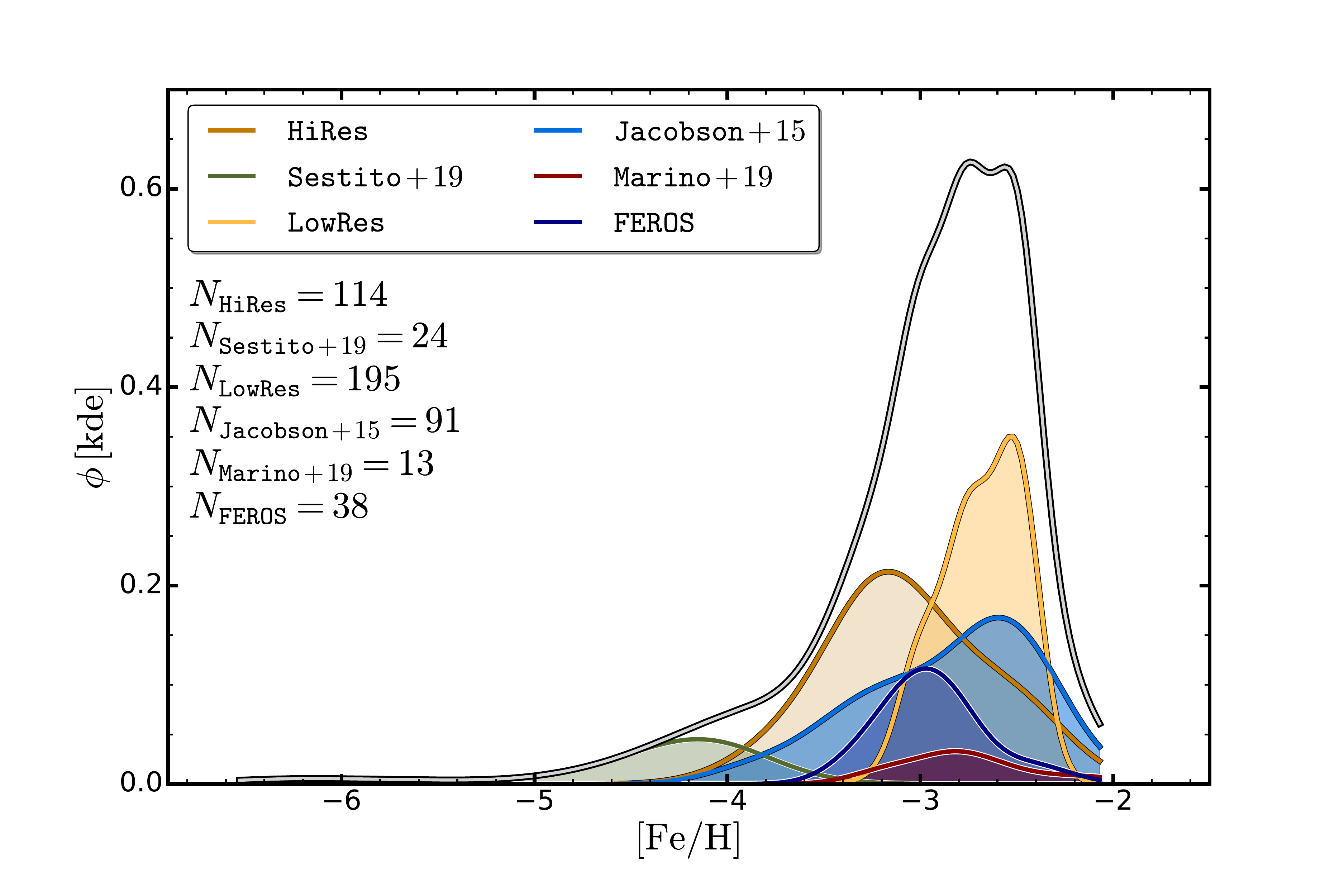}
		\caption{Metallicity distributions of the data sets analysed in this work. The six different subsamples are marked with red, orange, blue, dark-red, navy and green. %\DM{I'm not seeing red and green? More tan and grey} \GC{Changed the colours a bit. Better?}  
		%\GDC{will need to change the J+15 number to 89}
		The metallicity distribution of the total sample is shown with the grey-black solid line. Each density distribution ($\phi$) has been computed with a Gaussian kernel and renormalized with the total number of stars in the sample for a correct relative visualization.}
		\label{fig:metDistribution}
	\end{figure}

	Unless otherwise noted, the uncertainty in [Fe/H] values derived from high dispersion spectroscopy is taken as $\pm$0.10, while for the stars in the \texttt{LowRes} data set, the uncertainty is $\pm$0.3, and the values are quantized at 0.25 dex intervals. Figure~\ref{fig:metDistribution} then shows the metallicity distribution of each data set, computed using kernel density estimation with a Gaussian kernel and a bandwidth parameter of 0.5; the number of stars belonging to each set is reported in the top-left corner of the panel. Each distribution has been normalized by the number of stars in the sample.  
	Figure~\ref{fig:metDistribution} also shows the distribution for the total sample formed by summing the individual distributions.
	As is apparent, the sample spans a wide range in metallicity, with a peak around $\rm [Fe/H]\sim -2.8$, consistent with the observed metallicity distribution function of the full SkyMapper EMP sample discussed in \cite{dacosta2019}.

\begin{figure*}
		\centering
		\includegraphics[width=18cm, trim={2cm 0cm 2cm 0cm}, clip]{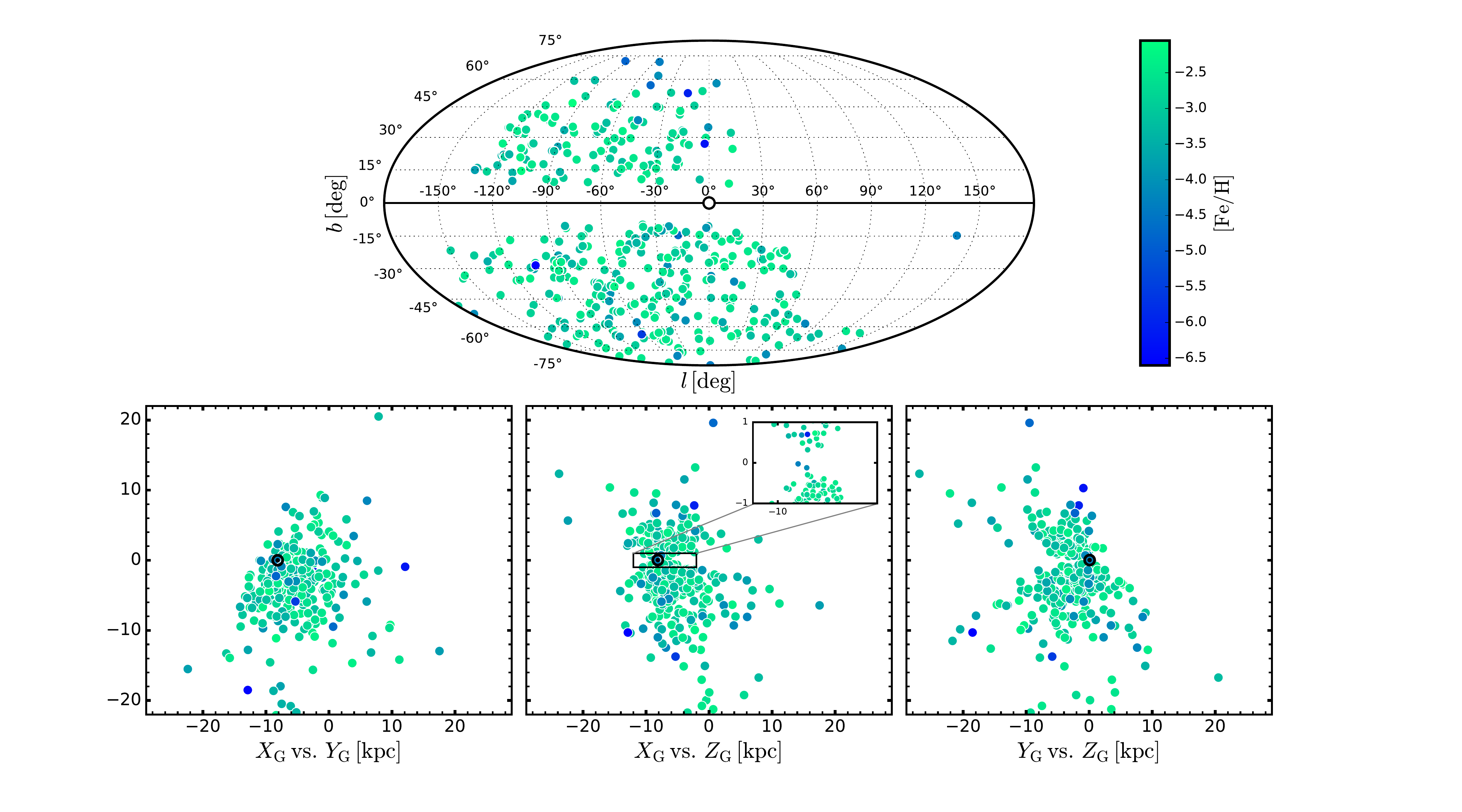}
		\caption{\textit{Top panel.} Mollweide projection of the analyzed stars in Galactic coordinates. Each star is colour coded according to its metallicity. \textit{Bottom panels.} Position of the analyzed stars in the Galactocentric Cartesian reference frame using the derived distances as discussed in section \ref{sub:dist}. The inset in the middle-bottom panel shows a zoom of the Galactic plane region. In each panel the first named quantity is for the x-axis and the second is for the y-axis.  The Sun, marked by the black circle, is at (--8.2, 0.0, 0.02) and the Galactic Centre is at the origin in this co-ordinate system. }
		
		\label{fig:mollweide}
	\end{figure*}

	Figure~\ref{fig:mollweide} shows the position of the analyzed stars, both in Galactic latitude and longitude and in the Cartesian Galactocentric reference frame, with each star colour-coded according to its metallicity. 
	Since $\sim\,90\%$ of the stars come from the SkyMapper survey, the data set is affected by the same selection biases as discussed in \citet{dacosta2019}. Specifically, the SkyMapper survey avoids regions of the sky with significant stellar crowding, while the selection process for candidates restricts the sample to stars with $E(B-V) < 0.25$ mag. The net result is a lack of candidates near the Galactic plane and in the Galactic Bulge \citep[see Figure~4 and Figure~14 in][]{dacosta2019} as is evident in the inset in the middle panel of Figure~\ref{fig:mollweide}. 
	Indeed, the majority of the stars lie inside the solar circle in the ($\rm{X}_{G},\rm{Y}_{G}$) plane, although at a variety of heights above and below the plane; the star nearest the Galactic Centre in the sample has a Galactocentric radius of $1.8\pm0.8$ kpc.  

\section{Deriving the kinematics of the sample}\label{sec:kin}

	To compute the orbit of a star the full 6-dimensional information for the position and velocity is needed. Specifically, we need right ascension
	($\alpha$), declination ($\delta$), distance from the Sun (d), proper motions in right ascension and declination ($\mu_{\rm \alpha}\cos\delta, \mu_{\rm \delta}$), and the heliocentric radial velocity $(v_{\rm r}$).  
	Gaia DR2 provides the optimal source for these parameters, noting that strictly Gaia provides a measurement of parallax, not distance, and that $v_{\rm r}$
	is only available for the brightest stars.
	%at least for the brightest stars as regards $v_{\rm r}$.   
	As recently discussed in \citet{bailerjones2018}, for example, simply inferring the distance from the parallax measurement alone can lead to unreliable results. To overcome this problem, \citet{bailerjones2018} combined parallax measurements with a realistic prior for the distance as a function of Galactic longitude and latitude, to generate distance estimates. 
	
	\citet{sestito2019a} introduced an alternate approach to determining distances that combines the exquisite astrometry and photometry provided by Gaia DR2 with theoretical isochrones in a Bayesian analysis to infer the distance, as well as the physical properties surface gravity $(\log\,g)$ and effective temperature $(T_{\rm eff})$. An advantage of their technique is that it allows the breaking of the potential degeneracy between dwarf and giant star distances at a fixed $T_{\rm eff}$. 
	In our case however, by deliberate choice of the colour-range used to define the underlying sample of low metallicity candidates in the SkyMapper EMP-survey \citep[see][]{dacosta2019}, our data set consists entirely of giants\footnote{This is verified by the $\log\,g$ values for our stars as determined from the high resolution spectra, where available, or from the spectrophotometric fits to the low-resolution spectra \citep[see][for details]{dacosta2019}.}, so that any dwarf/giant distance ambiguity does not arise.  It further allows us to exploit the effective temperatures and metallicities of our stars, which are known from either the high-resolution analyses or from the spectrophotometric fits to the low-resolution spectra,  to derive absolute magnitudes via the use of red giant branch (RGB) isochrones, particularly for those stars that lack a reliable parallax determination. This approach has the underlying assumption that all the stars lie on the RGB, whereas the distribution of temperatures and gravities in \citet{dacosta2019} suggests a small fraction (5--10\%) of the total sample are red horizontal branch or early-AGB stars.  Such stars are more luminous than RGB stars at the same effective temperature and thus the distance determinations based on the RGB locus will be smaller than the true distances.  While this will result in some individually incorrect orbital parameters, the overall results are unaffected given the dominance of RGB stars in the sample.
	
\begin{figure}
			\centering
			\includegraphics[width=8.9cm,, trim={0cm 0cm 0cm 0cm}, clip]{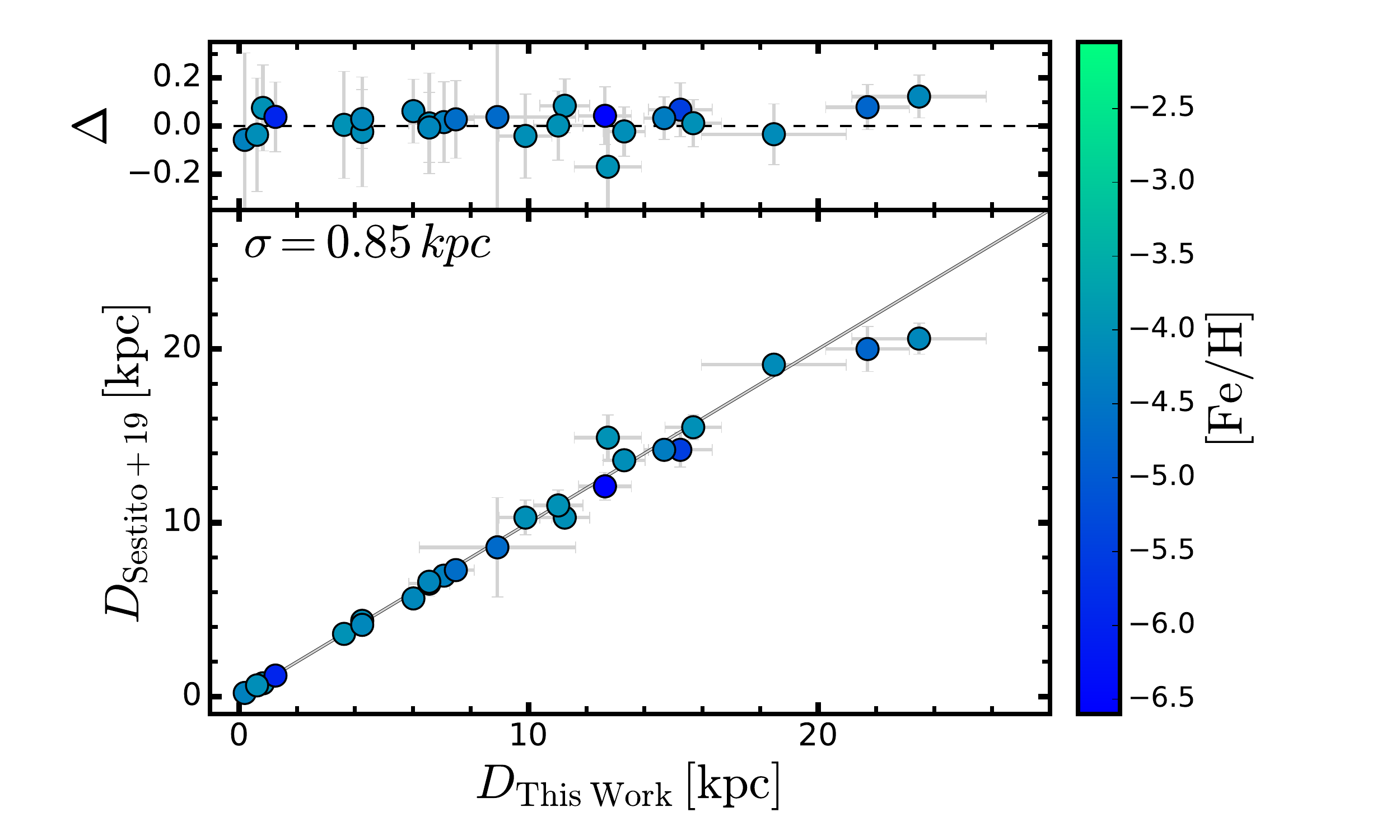}
			\caption{{\it Lower panel:} Comparison between the distance estimates in this work and those from \citet{sestito2019a}. The solid line represents the 1:1 relation between the two different estimates. {\it Upper panel:} the relative differences between our distances and those from \citet{sestito2019a} expressed as $( \Delta=(D_{\rm TW}-D_{\rm S+19})/D_{\rm TW})$. The subscript TW indicates the values from this work. Each star is colour-coded according to its metallicity, as shown by the colour bar. %The latter option would be consistent with Figure 4
			}
			\label{fig:compSestito}
		\end{figure}

	Details of our distance determinations are discussed in the next section, but when we compare our derived distances with those in \citet{sestito2019a} for the stars in common, we find excellent agreement.  This is illustrated in Figure~\ref{fig:compSestito}. 
	
	\subsection{Distance determination}\label{sub:dist}

		Our approach to determining distances for the stars in our sample is twofold. First, 
		for the stars with unreliable Gaia DR2 parallax determinations, which we take here as those with  $\sigma_{\rm \pi}/|\pi| \geq 0.15$, we adopted the following approach, which relies on the assumption that the stars in our sample, being very metal-poor, can be safely assumed to be old (age $\geq$ 10 Gyr)\footnote{At ages exceeding $\sim$10 Gyr, for a given isochrone set there is very little variation in absolute magnitude with age at fixed metallicity and $T_{\rm eff}$ on the RGB.}.  With this assumption we can then use the known $T_{\rm eff}$ and [Fe/H] values together with RGB isochrones of different metallicity to infer the absolute magnitudes and thus the distance.  
		Specifically, we have used a set of Yonsei-Yale RGB isochrones\footnote{These isochrones were adopted for consistency with the analyses in \citet{jacobson2015,marino2019} and in the \texttt{HiRes dataset} (Yong et al.\ (in preparation), where the isochrones were used to infer surface gravities.}
		($Y^2$, \citet{demarque2004}) for an age 12 Gyr, $[\alpha/{\rm Fe}] = +0.3$ and metallicities corresponding to [Fe/H] = --3.5, --2.5 and --1.9 to infer the V-band absolute magnitude $(M_{\rm V})$ for each star.  
		
		In practice, to find the absolute magnitude corresponding to a given star's metallicity and $T_{\rm eff}$, we interpolated in $M_{\rm V}$ across the isochrones at the $T_{\rm eff}$ value. Since the isochrones use visual magnitudes, we first calculated the appropriate $V$ magnitude for each star from the Gaia $G$ values using the coefficients provided by the Gaia documentation\footnote{\url{https://gea.esac.esa.int/archive/documentation/GDR2/Data_processing/chap_cu5pho/sec_cu5pho_calibr/ssec_cu5pho_PhotTransf.html}}.
		Reddening values from \citet{schlegel1998} were adopted, corrected according to the recipe in \citet{wolf2018}.
		For stars with metallicities between $-4.5$ and $-3.5$, the $(M_{\rm V})$ value is a linear extrapolation, while for the small number of stars with [Fe/H] $\leq-4.5$, which come primarily from the \citet{sestito2019a} sub-sample, the $M_{\rm V}$ inferred for [Fe/H] = $-4.5$ was used.
		The uncertainties in the distances were then determined by assuming an uncertainty of $100\,\rm K$ in \teff  and $0.1\,\rm dex$ in metallicity ($0.3\,\rm dex$ for stars in \texttt{LowRes} subsample) and then propagating these values into the distance determination.
		
\begin{figure}
			\centering
			\includegraphics[width=8.8cm, trim={0cm 0cm 0cm 0cm}, clip]{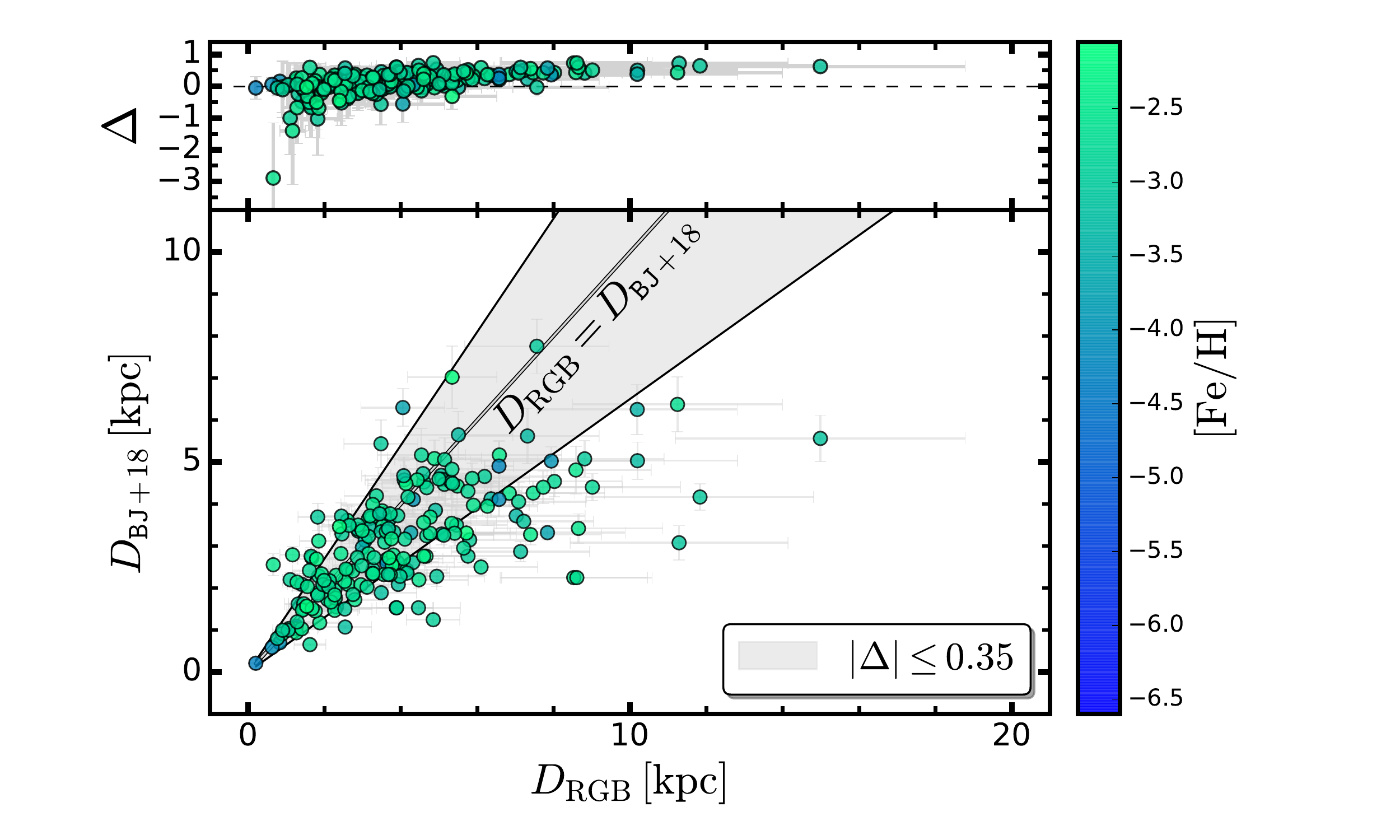}
			\caption{Comparison between the distance determined in this work and the distance inferred in \citet{bailerjones2018} for the 195 stars with $\sigma_\pi/|\pi|<0.15$. Each star is colour coded according to its metallicity, as shown in the right colour-bar. The grey shaded region within the black solid lines encloses stars with $|\Delta| \leq 0.35$, for which we adopted the Bailer-Jones distances. The top panel shows the relative differences between distances inferred through RGB isochrones and distances from \citet{bailerjones2018} $( \Delta=(D_{\rm RGB}-D_{\rm BJ+18})/D_{\rm RGB})$}
			\label{fig:distDet}
		\end{figure}

		Second, for the stars with nominally reliable Gaia DR2 parallax determinations, i.e., those with $\sigma_{\rm \pi}/|\pi| < 0.15$, we compared the \citet{bailerjones2018} distances with the distances inferred from the RGB isochrones.  This is shown in Fig.\ \ref{fig:distDet}.
		While most stars do scatter about the 1:1 line, there are sizeable differences between the two estimates for $\sim$25\% of the stars, most commonly with the RGB-based distance being larger than the \citet{bailerjones2018} value, indicating that the parallax may have been overestimated, or that the RGB-based distance is incorrect.  
		
		We have not sought to investigate the origin of the discrepancy for each individual case, noting that we include uncertainties in \teff and [Fe/H] when estimating the uncertainty in the RGB-based distance.  There is, however, a potential systematic uncertainty introduced by RGB isochrone based approach.  In particular, as discussed by \citet{joyce2018}, the location of theoretical RGBs in the Hertzsprung-Russell diagram is sensitive to the adopted value of the mixing length parameter $\alpha_{\rm MLT}$.  The value of $\alpha_{\rm MLT}$ employed in any particular isochrone set (e.g., $\alpha_{\rm MLT}$ = 1.7 for the $Y^2$ isochrones) is usually determined by requiring a fit to the solar values, but, as demonstrated in \citet{joyce2018}, at low metallicities the location of the RGB computed with a solar-calibrated $\alpha_{\rm MLT}$ is more luminous by $\sim$0.3 mag at a fixed $T_{\rm eff}$ than a comparison with globular cluster RGB observations would suggest: a $\sim$10\% smaller value of $\alpha_{\rm MLT}$ is required for consistency with the observations.  It is possible therefore that our RGB-based distances are systematically over-estimated, though the comparison shown in Fig.\ \ref{fig:distDet} suggests that it is not a major effect.  
		
		In practice %to maintain a consistent approach, [not sure what 'consistent approach actually meant, so have removed it]
		we have adopted the \citet{bailerjones2018} distance and its uncertainty whenever the absolute value of the relative difference $(\Delta = \frac{D_{\rm RGB}-D_{\rm BJ+18}}{D_{\rm RGB}})$ is smaller than 0.35. This is shown as the grey shaded region in Figure~\ref{fig:distDet}. For the remainder, i.e., for the stars outside the shaded area with $|\Delta|>0.35$, we adopted the distance inferred from RGB isochrones.
		Overall, this results in the use of the RGB isochrone distance for 357 stars, while for the remaining 118 the \citet{bailerjones2018} distance is 
		employed\footnote{We have checked that the kinematics for the stars where we have adopted the \citet{bailerjones2018} distance are not significantly altered if instead the RGB distance is assumed.  This is not surprising as for these stars the \citet{bailerjones2018} and RGB distances are consistent.}. 
		
        The largest (heliocentric) distance of the stars in our sample is the RGB-based distance of $\sim35$ kpc for the high-luminosity giant ($\log g$ $\approx$ 0.3) SMSS~J004037.55--515025.1, which has [Fe/H] = --3.83 and is in the \texttt{Jacobson+15} sub-sample.  The smallest is the \citet{bailerjones2018} distance of 0.21 kpc for the sub-giant BD+44~493 ($\log g$ $\approx$ 3.2 and [Fe/H] = --4.30) from the \texttt{Sestito+19} sub-sample.
		Overall, the median heliocentric distance for the entire sample is $\sim 5$~kpc, with a median $\sim 7$~kpc for RGB-based distances, and a median of $\sim 3$~kpc for \citet{bailerjones2018} distances.

	\subsection{Orbital properties}\label{sub:orb}
	
		To compute the orbital parameters we used the full 6-dimensional information on the position and velocity for each star. Gaia DR2 provides coordinates and proper motions, while the distances have been obtained as discussed in the previous section. Radial velocities come from the high-dispersion spectra, when available, and from Gaia DR2 for the \texttt{LowRes} sample.  
		
		We note that there are a number of the stars with radial velocities from the high-dispersion spectra that also have radial velocities from Gaia DR2, and this allows us to check for anything unusual or unexpected.  As mentioned in \S \ref{sec:data}, in this comparison process we discovered an anomaly in the \citet{jacobson2015} radial velocities for a particular Magellan/MIKE run. In that run 32 stars were observed of which 8 also have radial velocities from Gaia DR2.  The comparison for these 8 stars shows extreme disagreement for 7 stars, with values of the difference V$_{r}$(J+15) -- V$_{r}$(Gaia DR2) ranging from --400 to +415 $\rm km\,s^{-1}$.  We are at a loss to explain the origin of the disagreements\footnote{It is important to recall that Gaia DR2 velocities were not available at the time the \citet{jacobson2015} results were published, so the velocity anomalies would not have been apparent.} and have consequently excluded from the analysis the stars from this run that lack Gaia DR2 radial velocities, while using the Gaia DR2 radial velocities in the kinematic calculations for the remaining 7 stars that have [Fe/H] $\leq$ --2.0 dex.  We stress that such large disagreements are seen only for this one observing run in the \texttt{Jacobson+15} sample, the radial velocities from other runs are very consistent with Gaia DR2 values when available.  This is also the case for the stars in the \texttt{HiRes} and \texttt{FEROS} samples.  Overall, for the 41 stars with radial velocities from our high-dispersion spectra and from Gaia DR2, the velocities agree well with a mean difference, in the sense of our velocities minus Gaia, of 1.7 $\rm km\,s^{-1}$ and a standard deviation of 5.5 $\rm km\,s^{-1}$. This agreement indicates that any systematic uncertainties in the radial velocity determinations are very minor compared to other contributors to uncertainties in the orbit determinations.  We have always used the radial velocity and the corresponding uncertainty from the high dispersion spectra when available; the Gaia radial velocities and their uncertainties were utilized only when there was no alternative.

		The kinematics of our sample of metal-poor stars have been determined using the \texttt{GALPY}\footnote{\url{http://github.com/jobovy/galpy}} Python package \citep{bovy2015}.  The orbit of each star was obtained by direct integration backward and forward in time for 2\,Gyrs.This choice relies on the assumption that such a timescale is shorter than any significant variation in the Galactic potential
		
		We adopted the potential identified as the best candidate among the ones studied in \citet{mcmillan2017}\footnote{We note that our adopted potential is different from that used in \citet{sestito2019a}.}. Briefly, it consists of an axisymmetric model with a bulge, thin, thick and gaseous disks, and a Navarro-Frenk-White
		\citep{NFW} dark matter halo. %The Python implementation is publicly available on {\it GitHub}\footnote{e.g. \url{https://github.com/jamesmlane}}. Comment: Steph M says this isn't necessary.
		The heliocentric distances derived in the previous section have been converted to distances from the Galactic Centre (GC) in the \texttt{GALPY} routine, specifying the galactocentric position of the Sun as $(X, Y, Z)=(-8.21, 0, 0.0208)\,\rm kpc$, and its circular speed as $v_0=232.8\, \rm km\,s^{-1}$.  Both quantities are taken from \citet{mcmillan2017}.

		For each star we determined the apogalacticon and perigalacticon $(D_{\rm apo}, \, D_{\rm peri})$ of the orbit, the maximum vertical excursion from the Galactic plane $(Z_{\rm max})$, the eccentricity $\left(e=\frac{D_{\rm apo}-D_{\rm peri}}{D_{\rm apo}+D_{\rm peri}}\right)$, the energy $(E)$, the three actions $(J_{\rm R},\, J_{\rm \phi},\, J_{\rm Z})$\footnote{See \citet{binney12} for a description of these variables. In particular, the azimuthal action $J_{\rm \phi}$ corresponds to the vertical angular momentum $L_{\rm Z}$ for an axisymmetric potential, as is the case here.  
		In the following we will therefore refer to the azimuthal action in place of the vertical angular momentum.  We also adopted the St{\"a}ckel fudge method to calculate the actions as implemented in \texttt{GALPY}. } and the velocity components $U,\, V,\, W$ in the frame of the local standard of rest (LSR). As have others \citep[e.g.][]{myeong2018}, we emphasize that action space is the ideal plane in which to evaluate large samples of MW stars to identify and study possible sub-structures and debris from accretion events. The reason is that the actions are nearly conserved under the hypothesis that the potential is smoothly evolving \citep{binney1984}.
		
        The uncertainties associated with the derived orbital parameters are determined by sampling the Probability Distribution Functions (PDFs) of the observed values. In particular, we drew 500 random realizations of the distance and velocity components and, for each realization, recomputed the orbital parameters
		assuming Gaussian distributions with means and dispersions equal to the observed values and their uncertainties. 
        In particular, the uncertainties in the two proper motion components $(\mu_{\rm \alpha}\cos\delta, \, \mu_{\rm \delta})$ were drawn from a bivariate Gaussian taking into consideration the full covariance as defined in Equation~\ref{eqn:cov}, following the Gaia DR2 documentation.
		\begin{equation}
			cov =   
			\begin{pmatrix}
    			\sigma^2_{\rm \mu_{\rm\alpha }} & \sigma_{\rm \mu_{\rm\alpha }}\cdot\sigma_{\rm \mu_{\rm\delta }}\cdot corr(\mu_{\rm \alpha}, \mu_{\rm \alpha})  \\
    			\sigma_{\rm \mu_{\rm\alpha }}\cdot\sigma_{\rm \mu_{\rm\delta }}\cdot corr(\mu_{\rm \alpha}, \mu_{\rm \alpha}) & \sigma^2_{\rm \mu_{\rm\delta}}
  			\end{pmatrix}
  			\label{eqn:cov}			
		\end{equation}
        The uncertainties on the orbital parameters have then been determined by propagating the $16^{\rm th}$ and $84^{\rm th}$ percentiles of the resulting parameter distributions.

        As examples, we consider the two most iron-poor stars known: SMSS~J031300.36--670839.3 \citep{keller2014,bessell2015} and SMSS~J160540.18--144323.1 \citep{nordlander2019}.  For the former we find an ``outer-halo'' orbit with $e = 0.70 \pm 0.05\,, D_{\rm peri} = 6.5 \pm 2.0,\, D_{\rm apo} = 36.6 \pm 9.8$  and $|Z_{\rm max}| = 34.2 \pm 9.2~{\rm kpc}$.  These parameters are in good agreement with those listed in \citet{sestito2019a}.  For the latter star, however, we determine an extreme ``outer-halo'' orbit that may in fact be unbound as the derived energy $E$ is close to zero.  The inferred parameters are $e = 0.93,\, D_{\rm peri} = 6.5 \pm 2.0, D_{\rm apo} \approx 423$ and $|Z_{\rm max}| \approx 327~{\rm kpc}$; the latter two quantities are quite uncertain.
		
        As discussed in detail in \S \ref{sub:escapers}, SMSS~J160540.18--144323.1 is, in fact, one of a small number of stars (30 out of 475) for which we find apparent apogalacticon distances larger than the Milky Way virial radius, i.e., larger than $\sim$250 kpc. For such stars, a substantial fraction of the 500 random realizations %(specifically larger than 42.6\%) \DM{What's the significance of this number?} \GC{It's the percentage of random realizations which resulted in unbound orbits. It means that Galpy found and infintite apocenter in at least 42.6\% of the 500 random realizations of stars with apocenter larger than 250\,kpc. Was that the question?} 
		resulted in unbound orbits (i.e., $D_{\rm apo}=\infty$), thus potentially biasing both the medians and the uncertainties derived from the orbital parameter distributions. The uncertainties for these specific stars are considered in more detail in \S \ref{sub:escapers}.
		
		As an independent check on the uncertainties and on the role of the adopted potential, we can compare our orbit parameters with those listed in 
		\citet[][see their Table 4]{sestito2019a} for the 24 Ultra Metal-Poor stars in common.  The agreement is generally excellent.  Specifically, defining $\Delta$ as the difference between our values and those of \citet{sestito2019a} normalized by our values, then for the 24 stars we find median $\Delta$ values of 0.04, 0.03, 0.03 and 0.08 for $D_{\rm apo}$, $D_{\rm peri}$, $J_{\rm \phi}$ and $E$, respectively, noting that for the energy comparison we have taken into account the different solar energy used here to that in the \citet{sestito2019a} study.

\section{Results}\label{sec:results}

	\begin{figure*}
	 	\centering
	 	\includegraphics[width=0.95\textwidth, trim={5cm 0cm 2cm 0cm}, clip]{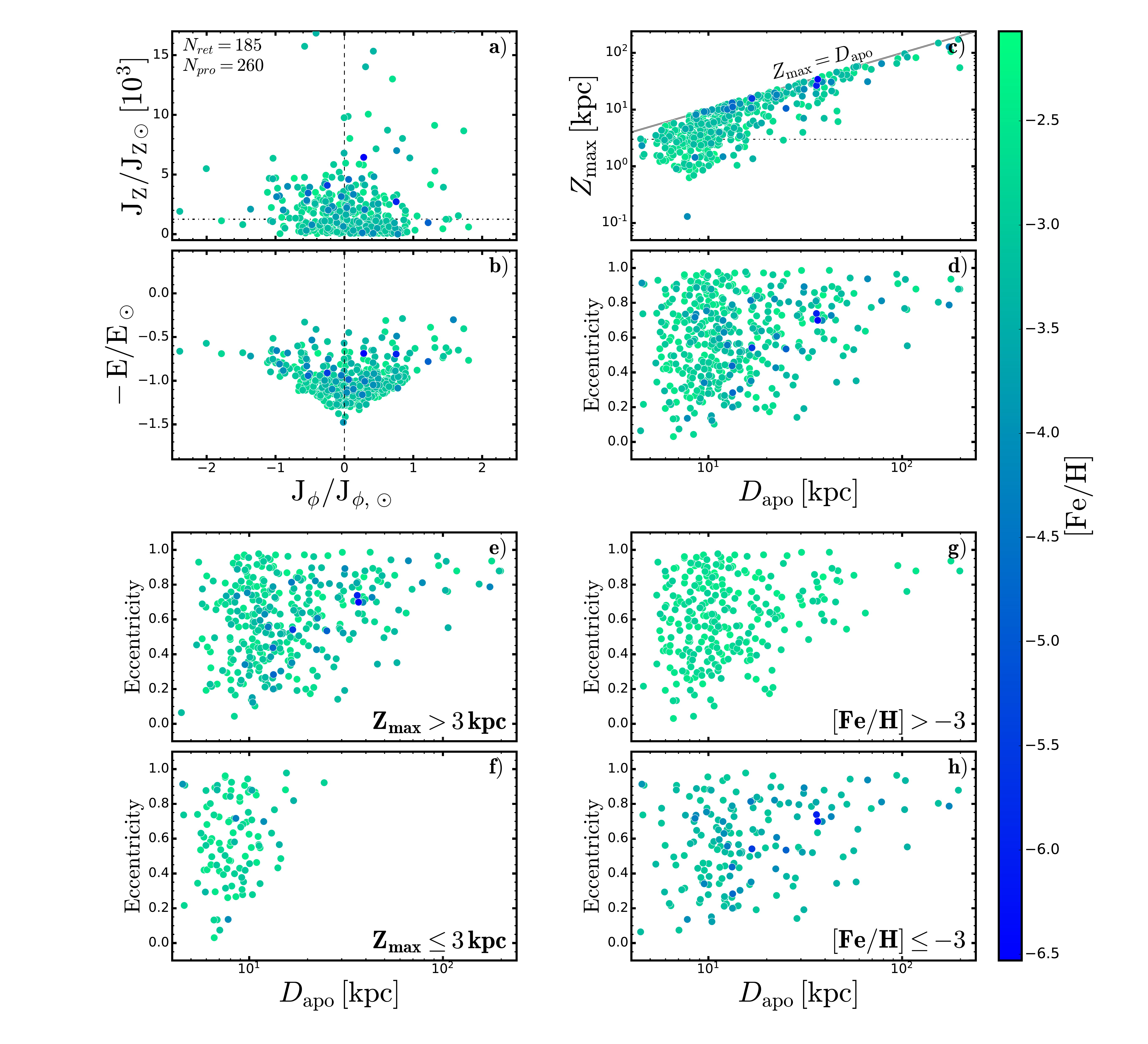}
	 	\caption{Orbital parameters for the stars with $D_{\rm apo} \leq $ 250\,kpc.  \textit{Panels a) and b):} Vertical action $(J_{\rm Z}\,\rm [km\,s^{-1}])$ and energy $(E\,\rm [km^2\,s^{-2}])$ as a function of the azimuthal action (i.e. the vertical component of the angular momentum, $J_{\rm \phi}\,\rm [km\,s^{-1}]$). All quantities have been normalized by the solar values. The horizontal dashed-dotted line in panel a) indicates $J_{\rm Z}/J_{\rm Z, \odot}=1.25\times10^3$.  
	 	\textit{Panels c) and d):} Maximum altitude from the MW plane $(|Z_{\rm max}|\,\rm [kpc])$ and eccentricity plotted as a function of the apogalactic $(D_{\rm apo}\,\rm[kpc])$ distance.  Note that since $|Z_{\rm max}|$ cannot exceed $D_{\rm apo}$, the region above the 1:1 line in panel c) is forbidden. The horizontal dashed-dotted line in panel c) marks $Z_{\rm max}=3\,\rm kpc$. 
	 	\textit{Panels e) and f):} as for panels c) and d) but split by $|Z_{\rm max}|$.  \textit{Panels g) and h):} as for panels c) and d) but split by metallicity [Fe/H].  %Typical errors are shown in the lower right of each panel.
	 	}
	 	
	 	\label{fig:allsamples}
	\end{figure*} 
	
	The physical properties and the computed orbital parameters of the first 10 stars are listed in Tables~\ref{tab:phys_properties} and \ref{tab:orb} while the complete tables are available with the online supplementary material.  For Table~\ref{tab:phys_properties} the columns are, respectively, an index number, the Gaia DR2 and SkyMapper or other IDs, the on-sky location in degrees, the parallax and its uncertainty from Gaia DR2, the adopted distance and its uncertainty, a flag indicating whether the distance is from the RGB isochrones (value=0), or from \citet{bailerjones2018} (value=1), the proper motions from Gaia DR2 and their uncertainties, 
	the heliocentric radial velocity and its uncertainty, $\log\,T_{\rm eff}$ and its uncertainty, the abundance [Fe/H], the reddening, and the data set from which the star originates. 
	Similarly for Table~\ref{tab:orb}, the columns are the index number (as for Table~\ref{tab:phys_properties}), the eccentricity and its uncertainty, the apo- and peri-galactic distances, the maximum deviation from the Galactic plane,
	the actions $(J_{\rm R},\, J_{\rm \phi},\, J_{\rm Z})$, the energy, and the $U$, $V$ and $W$ velocity components in the LSR frame.  %\GDC{The full electronic versions of both Tables will need to be updated but the versions in the text are OK}.
	
	Figure~\ref{fig:allsamples} shows the inferred orbital parameters for all the stars with $D_{\rm apo} \leq $ 250 kpc, with each star colour-coded according to its metallicity, as shown in the right colour-bar. %The typical uncertainties are shown in the bottom-right corner of each panel.  
	In particular, panels a) and b) show the vertical action $(J_{\rm Z}\,[\rm kpc\,\rm km\,s^{-1}])$, indicative of the vertical excursion of the star, and the orbital energy $(E\,\rm [km^2\,s^{-2}])$\footnote{The energy is multiplied by $-1$ to maintain the canonical ``V''-shape.}, as a function of the azimuthal action $(J_{\rm \phi}\,\rm [\rm kpc\,\rm km\,s^{-1}])$\footnote{The azimuthal action corresponds to the vertical angular momentum $L_{\rm Z}$ for an axisymmetric potential as is used here.}. The quantities have been normalized by the solar values computed for the \texttt{McMillan2017} potential employed here: $J_{\rm \phi, \odot}=2014.24\,\rm kpc\,\rm km\,s^{-1}$, $J_{\rm Z, \odot}=0.302\,\rm \rm kpc\,\rm km\,s^{-1}$ and $E_{\rm \odot}=-153507.15\,\rm km^2\,s^{-2}$.  We note that if we adopt the \texttt{MWPotential2014} employed by \citet{sestito2019a}, and include the increased dark matter halo mass, we obtain solar values similar to those in that work.
	Retrograde orbits are characterised by a negative value of $J_{\rm \phi}$, while prograde orbits have a positive $J_{\rm \phi}$. We find that overall $\sim42\%\, (185/445)$ of our stars with $D_{\rm apo} \leq $ 250 kpc exhibit retrograde orbits, and note that the selection of the stars for inclusion in our sample should not have any bias as regards prograde or retrograde orbits.
	
	Regarding the uncertainties in the derived orbital quantities, these are listed for each individual star in Table \ref{tab:orb}, but as examples, we find that for stars with $D_{\rm apo} \leq $ 20 kpc, the median errors in 
	$D_{\rm apo}$, $D_{\rm peri}$, $|Z_{\rm max}|$ and $e$ are 0.9 kpc, 0.6 kpc, 1.0 kpc and 0.08, respectively.  These increase to 11.6 kpc, 1.7 kpc, 7.7 kpc and 0.08 for stars with 
	$20 \leq D_{\rm apo} \leq $ 50 kpc, respectively, and to 56.4 kpc, 2.7 kpc, 52.0 kpc and 0.09 for stars with $50 \leq D_{\rm apo} \leq $ 250 kpc.
	
	In panels c) and d) we show the maximum height $(|Z_{\rm max}|\, \rm [kpc])$ and eccentricity $e$ as function of the 
	apogalatic distance $(D_{\rm apo}\,\rm [kpc])$. 
	A preliminary inspection of panels a) and c) reveals that, despite the low metallicity of the stars in the sample, we detect a significant number of stars with small vertical excursion, in agreement with \citet{sestito2019a,sestito2020} and \citet{dimatteo2019}. In particular, if we follow \citet{sestito2020} and
	adopt $J_{\rm Z}/J_{Zsun} < 1.25 \times 10^{3}$, shown as the dotted horizontal line in panel a), to characterize orbits that are confined to the disk, then $\sim$50\% of our sample meets this definition. 
	Similarly, if we follow \citet{sestito2019a} in using $|Z_{\rm max}|$ = $3\,\rm kpc$ (horizontal dashed-dotted line in panel c) of Figure~\ref{fig:allsamples}) to discriminate between ``disk-like'' and ``halo-like'' orbits, we find that 102, or $\sim$21\%, of the stars in our sample meet this criterion, i.e., have orbits that do not deviate far from the Galactic plane.  
	%We note that beyond $D_{\rm apo}$ $\approx$ 50--60~kpc in panel c), there is an apparent dearth of stars with low values of $|Z_{\rm max}|$.  However, we cannot rule out that this is a  result of a selection effect in the SkyMapper EMP survey that mitigates against finding stars with smaller $|Z_{\rm max}|$ at large $D_{\rm apo}$.
	Further, panel d) suggests that, while stars with $D_{\rm apo}\lesssim25\,\rm kpc$ have an approximately uniform distribution in eccentricity, highly eccentric ($e\gtrsim 0.5$) orbits are favoured for stars with $D_{\rm apo}\gtrsim25\,\rm kpc$, while panels c) and f) show that there is an apparent dearth of stars with low values of $|Z_{\rm max}|$ beyond $D_{\rm apo}$ $\approx$ 30~kpc. These apparent effects are most probably a consequence of the criteria adopted to select SkyMapper EMP candidates, as stars with low $|Z_{\rm max}|$
	%eccentricity, i.e.\ approximately circular orbits,
	and large $D_{\rm apo}$ aren't likely to meet the apparent magnitude cut that underlies the sample ($g_{\rm skymapper} < 16$ for the \texttt{HiRes} stars and $g_{\rm skymapper} < 13.7$ for the \texttt{LowRes} stars). In particular, the bottom-middle and bottom-right panels of Figure \ref{fig:mollweide} show that stars with $|Z|$ $\leq$ 3~kpc and Galactocentric distances beyond 10-15~kpc are rare in our sample.
	
	The two bottom left panels again show the eccentricity versus the apogalactic distance, but separately for stars with $|Z_{\rm max}|$ in excess of 3 kpc (panel e) and those with $|Z_{\rm max}|$ less than this value (panel f). Similarly, the two bottom right panels also show eccentricity versus the apogalactic distance but this time the sample is split by metallicity: stars with $\rm[Fe/H]> -3$ are shown in panel g) while the more metal-poor stars are shown in panel h). The similarity of panels g) and h) show that there is no obvious dependence of the kinematics on metallicity, at least for this sample of metal-poor stars.
	%These panels show that for the metal-poor stars in our sample there is little dependence of the kinematics on metallicity.
	
	\begin{figure}
        \centering
        \includegraphics[width=9cm, trim={1cm 0cm 0cm 1cm}, clip]{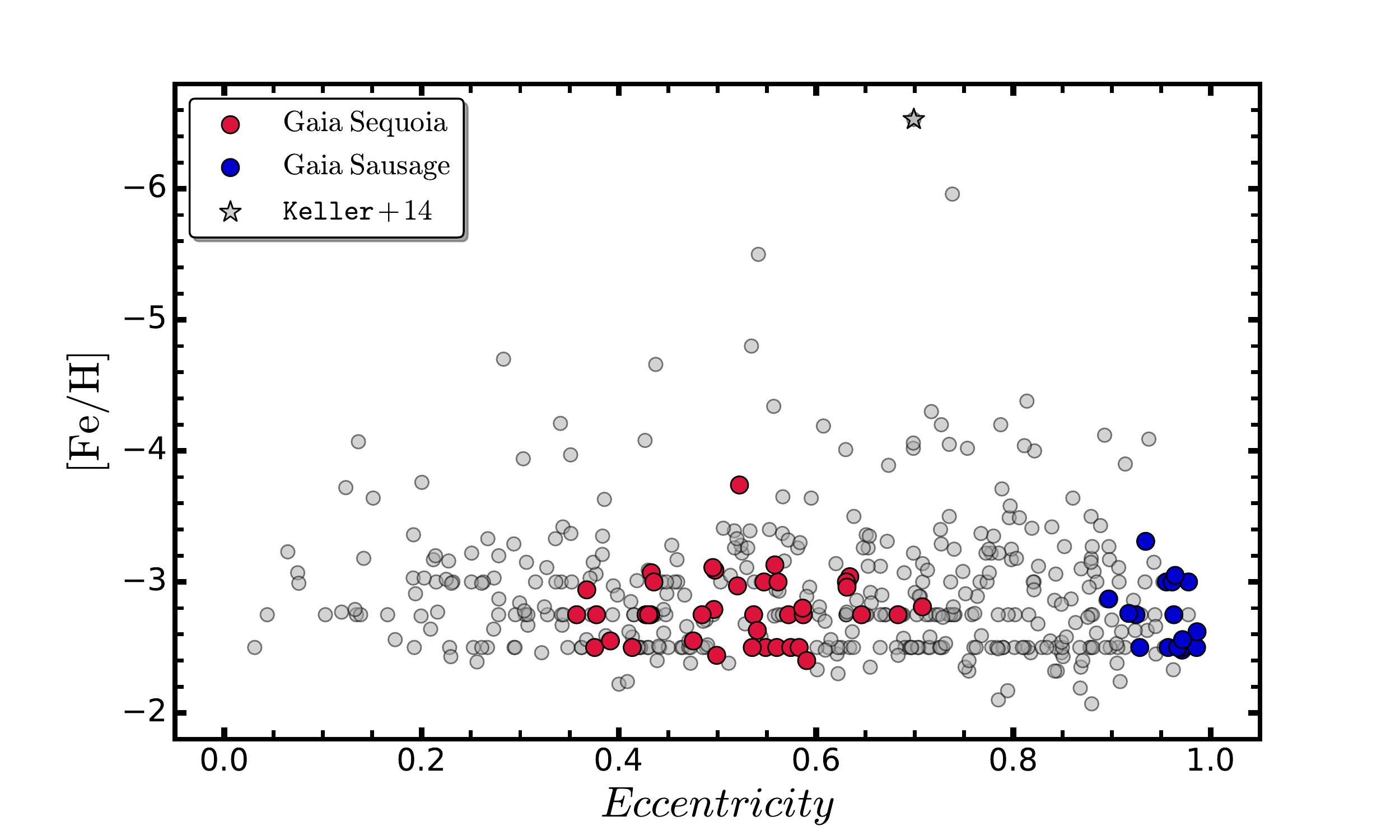}
        \caption{[Fe/H] vs. $e$ for the stars with $D_{\rm apo}<250\,\rm kpc$. \textit{Gaia Sausage} and \textit{Gaia Sequoia} candidates are shown with blue and red circles, respectively. The Keller star \citep{keller2014, nordlander2017} is shown as a grey star indicating the upper limit on the abundance.}
        \label{fig:eccvszmax}
	\end{figure}
	
	To more clearly illustrate this point, we show in Figure~\ref{fig:eccvszmax} a plot of [Fe/H] against $e$, the orbital eccentricity. Diagrams of this nature have long played an important role in discussions of the formation of the Galaxy.  For example, in their classic paper, \citet{ELS} argued on the basis of an apparent correlation between ultra-violet excess (an indicator of [Fe/H]) and orbital eccentricity, that the proto-Galaxy collapsed rapidly to a planar structure with a timescale of only a few $\times 10^{8}$ years.  Specifically, in their sample of stars, those with [Fe/H] less than --1.5, approximately, all had $e \geq 0.6$ \citep{ELS}.  \citet{norris1985} challenged the rapid collapse interpretation arguing that the lack of low-$e$ metal-poor stars was a result of a kinematic bias in the selection of the \citet{ELS} sample.  Instead, using a sample selected without any kinematic bias, \citet{norris1985} showed that metal-poor stars with relatively low orbital eccentricities exist, a population they identified as a metal-weak component of the thick-disk.  The \citet{norris1985} result was confirmed and strengthened by \citet[][see their Fig.\ 10]{beers2014} who showed that for stars with [Fe/H] $\leq$ --1.5 there is no correlation between orbital eccentricity and metallicity: stars can be found with $e$ values between $\sim$0.1 and 1.  Our results in Figure~\ref{fig:eccvszmax} extend the lack of any correlation to substantially lower metallicities than those in \citet{beers2014}, where there were only a few stars at or below [Fe/H] = --2.5 and none below --3.0 dex. We discuss the implications of the existence of extremely metal-poor stars with low eccentricities (and low $|Z_{\rm max}|$) in \S \ref{sub:disk}.  Figure \ref{fig:eccvszmax} also shows the location of candidate members of the \textit{Gaia Sausage} and \textit{Gaia Sequoia} accretion events.  The identification and properties of these stars are discussed in detail in \S \ref{sub:gsge}.
	
	Finally, as noted above, we find that 30 stars from the full sample have apparent $D_{\rm apo}$ values larger than 250~kpc, i.e., larger than the virial radius of the Milky Way.  The majority of these stars possess energies that are consistent with, or larger than, zero and they likely have unbound orbits.  These stars will be discussed in more detail in \S \ref{sub:escapers} but we note again that they are not plotted in the panels of Fig.\ \ref{fig:allsamples} or in Fig.~\ref{fig:eccvszmax}.

	\begin{table*}
		    \caption{Observed properties of the first ten stars in our sample.  The columns are: a numeral index, Gaia DR2 and SkyMapper or other IDs, coordinates, parallax and uncertainty, distance and uncertainty, a flag for distance method (0=RGB interpolation, 1=\citet{bailerjones2018}), proper motions and uncertainties, radial velocity and uncertainty, $\log T_{\rm eff}$ and uncertainty, [Fe/H], E(B-V) and origin data set as discussed in Section~\ref{sec:data}. The complete table is available electronically.}
	\centering
	    \begin{adjustbox}{width=1\textwidth}

        \begin{tabular}{lrrrrrrrrrrrrrrrrrrrr}
            \hline
            \hline
            Index & Gaia DR2 & SMSS J & $\alpha$ & $\delta$ & $\pi$ & $\sigma_{\rm \pi}$ & $D$ & $\sigma_{\rm D}$ & \texttt{FLAG} & $\mu_{\rm \alpha}$ & $\mu_{\rm \delta}$ & $\sigma_{\rm \mu_{\rm \alpha}}$ & $\sigma_{\rm \mu_{\rm \delta}}$ &  $v_{\rm r}$ & $\sigma_{\rm v_{\rm r}}$ & $log\,T_{\rm eff}$ & $\sigma_{\rm log\,T_{\rm eff}}$ &$\rm [Fe/H]$ & $\rm E(B-V)$ & $\rm Data set_{\rm id}$ \\
            
             & &  & $\rm deg$ & $\rm deg$ & $\rm mas$ & $\rm mas$ & $\rm kpc$ & $\rm kpc$ & & $\rm mas/yr$ & $\rm mas/yr$ & $\rm mas/yr$ & $\rm mas/yr$ & $\rm km\,s^{-1}$ & $\rm km\,s^{-1}$ & $\rm K$ & $\rm K$ & $\rm dex$ & \rm mag &  \\
            \hline
1 & 2398202677437168384 &  230525.31-213807.0 & 346.3555462 & -21.6353089 & 0.272750 & 0.049314 &   2.46 &    0.65 &       0 & -1.142 & -15.056 & 0.066 &  0.067  &  -15.7 & 0.4 & 3.708 & 0.009 & -3.26 & 0.027 &  \texttt{HiRes} \\
& & & & & & & & & & & & & & & & & & & & \\
2 & 2406023396270909440 &  232121.57-160505.4 & 350.3399235 & -16.0848819 & 0.418462 & 0.036657 &   1.10 &    0.20 &       0 & 17.161 &   3.631 & 0.069 &  0.054  &  -39.1 & 1.0 & 3.736 & 0.008 & -2.87 & 0.022 &  \texttt{HiRes} \\
& & & & & & & & & & & & & & & & & & & & \\
3 & 2541284393302759296 &  001604.23-024105.0 &   4.0177235 &  -2.6848020 & 0.282214 & 0.046544 &   2.98 &    0.78 &       0 & 13.561 &  -9.876 & 0.093 &  0.055  &   49.3 & 1.2 & 3.705 & 0.009 & -3.14 & 0.031 &  \texttt{HiRes} \\
& & & & & & & & & & & & & & & & & & & & \\
4 & 2623363791014198656 &  224145.62-064643.0 & 340.4401074 &  -6.7786758 & 0.030874 & 0.031151 &  12.07 &    3.06 &       0 &  1.853 &  -2.861 & 0.053 &  0.048  & -201.6 & 5.0 & 3.681 & 0.009 & -3.16 & 0.029 &  \texttt{HiRes} \\
& & & & & & & & & & & & & & & & & & & & \\
5 & 2666382767566459264 &  214716.16-081546.9 & 326.8173947 &  -8.2630725 & 0.499372 & 0.031415 &   3.48 &    0.93 &       0 &  1.497 & -37.651 & 0.057 &  0.049  &  -12.3 & 0.3 & 3.708 & 0.009 & -3.17 & 0.037 &  \texttt{HiRes} \\
& & & & & & & & & & & & & & & & & & & & \\
6 & 2909324470226028800 &  053721.56-244251.5 &  84.3398617 & -24.7143189 & 0.016537 & 0.027294 &   9.91 &    2.69 &       0 &  2.367 &   0.329 & 0.036 &  0.047  &  231.2 & 5.8 & 3.710 & 0.008 & -3.50 & 0.021 &  \texttt{HiRes} \\
& & & & & & & & & & & & & & & & & & & & \\
7 & 3064362275530429312 &  081627.99-055913.3 & 124.1166115 &  -5.9870501 & 0.176978 & 0.028682 &   7.47 &    1.92 &       0 & -0.403 &  -1.921 & 0.048 &  0.032  &  159.8 & 4.0 & 3.688 & 0.009 & -3.37 & 0.063 &  \texttt{HiRes} \\
& & & & & & & & & & & & & & & & & & & & \\
8 & 3064545859613457536 &  081112.13-054237.7 & 122.8005492 &  -5.7104991 & 0.098921 & 0.047035 &  21.76 &    5.71 &       0 &  0.245 &  -2.879 & 0.075 &  0.066  &  121.0 & 3.0 & 3.686 & 0.009 & -3.74 & 0.038 &  \texttt{HiRes} \\
& & & & & & & & & & & & & & & & & & & & \\
9 & 3458991567268745728 &  120218.07-400934.9 & 180.5752523 & -40.1597114 & 0.266276 & 0.035374 &   3.29 &    0.39 &       1 & 11.765 &  -2.682 & 0.040 &  0.029  &  -17.6 & 0.4 & 3.746 & 0.008 & -2.89 & 0.090 &  \texttt{HiRes} \\
& & & & & & & & & & & & & & & & & & & & \\
10 & 3473880535256883328 &  120638.24-291441.1 & 181.6593108 & -29.2447637 & 0.257289 & 0.024364 &   3.41 &    0.28 &       1 & -0.576 &  -2.246 & 0.031 &  0.016 &   58.1 & 1.5 & 3.708 & 0.009 & -3.06 & 0.052 &  \texttt{HiRes} \\
& & & & & & & & & & & & & & & & & & & & \\
            \hline
        \end{tabular}
        \end{adjustbox}
	    \label{tab:phys_properties}
	\end{table*}

	\begin{table*}
	        \caption{Derived orbital properties for the first ten stars in our sample. The columns are: numeral index; eccentricity; apo- and peri-perigalacticon distances; maximum height; orbital actions; orbital energy; $U$, $V$, $W$ velocities and orbit type (Halo, Disk, Sequoia, Sausage, Unbound).  Each numerical quantity is followed by upper and lower uncertainties. The complete table is available electronically.}
	\centering
	    \begin{adjustbox}{width=1\textwidth}	
        \begin{tabular}{lrrrrrrrrrrrc}

            \hline
            \hline
            index &    $\rm Eccentricity$ &  $D_{\rm apo}$ &  $D_{\rm peri}$ &  $Z_{\rm max}$ &   $J_{\rm R}$ &  $J_{\rm \phi}$ &  $J_{\rm Z}$ & $\rm Energy$ & $U$ & $V$ & $W$ & Orbit type \\
             &   &  $\rm kpc$ &  $\rm kpc$ &  $\rm kpc$ &   $\rm kpc\cdot km\,s^{-1}$ &  $\rm kpc\cdot km\,s^{-1}$ &  $\rm kpc\cdot km\,s^{-1}$ & $\rm kpc\cdot km^2\,s^{-2}$ & $\rm km\,s^{-1}$ & $\rm km\,s^{-1}$ & $\rm km\,s^{-1}$ & \\
            \hline
 1 & 0.58 $_{-0.20}^{+0.17}$ &  8.4   $_{-0.2}^{+0.3}$ &  2.2  $_{-0.9}^{+1.4}$ &  2.4  $_{-0.8}^{+1.0}$ &  295.5   $_{-138.6}^{+137.9}$ &   690.9   $_{-287.7}^{+360.1}$ &   96.8   $_{-45.0}^{+55.2}$ &  -176530 $_{-634}^{+2662}$ &   86.7 $_{-25.4}^{+22.1}$ & -147.8 $_{-42.4} ^{+48.3}$ &  3.9  $_{-4.8}^{+5.6}$ &     Disk \\
 & & & & & & & & & & & & \\ 
 2 & 0.28 $_{-0.06}^{+0.05}$ & 10.4   $_{-0.6}^{+0.5}$ &  5.9  $_{-0.3}^{+0.5}$ &  1.4  $_{-0.3}^{+0.3}$ &  104.6    $_{-43.7}^{+44.7}$ &  1698.7    $_{-44.4}^{+51.1}$ &   33.6    $_{-8.9}^{+9.0}$ &  -156666  $_{-697}^{+938}$ &  -83.8 $_{-15.6}^{+19.3}$ &  -16.0 $_{-2.8}^{+3.4}$ &   16.6  $_{-4.9}^{+6.0}$ &     Disk \\
 & & & & & & & & & & & & \\ 
 3 & 0.71 $_{-0.28}^{+0.17}$ & 11.2   $_{-1.3}^{+1.7}$ &  1.9  $_{-1.1}^{+2.0}$ &  6.3  $_{-2.6}^{+2.7}$ &  540.3   $_{-319.5}^{+227.7}$ &   456.2   $_{-522.2}^{+582.2}$ &  298.2  $_{-128.8}^{+127.1}$ &  -162656  $_{-2240}^{+6212}$ &  -91.6 $_{-26.4}^{+30.5}$ & -165.6 $_{-52.5}^{+62.1}$ & -121.1 $_{-22.9}^{+25.5}$ &     Halo \\
 & & & & & & & & & & & & \\ 
 4 & 0.57 $_{-0.08}^{+0.12}$ & 12.7   $_{-2.0}^{+3.0}$ &  3.5  $_{-1.1}^{+1.2}$ & 10.8  $_{-0.5}^{+1.8}$ &  460.7    $_{-87.2}^{+84.5}$ &  -494.8   $_{-292.1}^{+319.1}$ &  715.3   $_{-79.9}^{+204.8}$ &  -154330  $_{-7291}^{+10450}$ &  -63.8  $_{-6.4}^{+4.9}$ & -253.2 $_{-44.3}^{+39.1}$ &   58.4 $_{-33.3}^{+28.2}$ &     Halo \\
 & & & & & & & & & & & & \\ 
 5 & 0.80 $_{-0.17}^{+0.14}$ & 51.8  $_{-23.2}^{+98.8}$ &  5.8  $_{-2.5}^{+0.4}$ & 41.0 $_{-18.7}^{+81.7}$ & 2937.5  $_{-2646.9}^{+19134394.1}$ & -1367.7   $_{-431.8}^{+748.7}$ & 1187.9  $_{-563.7}^{+661.6}$ & -92378  $_{-63565}^{+86287}$ &  245.3 $_{-75.1}^{+65.1}$ & -515.5 $_{-141.5}^{+163.7}$ & -225.7 $_{-66.0}^{+75.8}$ &     Halo \\
 & & & & & & & & & & & & \\ 
 6 & 0.64 $_{-0.03}^{+0.02}$ & 18.8   $_{-2.9}^{+3.4}$ &  4.1  $_{-0.6}^{+1.2}$ &  5.8  $_{-2.7}^{+6.4}$ &  764.6   $_{-104.6}^{+72.7}$ &  1488.5   $_{-186.2}^{+149.5}$ &  153.8   $_{-78.2}^{+311.1}$ &  -136232  $_{-8547}^{+9007}$ & -141.8 $_{-5.4}^{+6.1}$ & -194.3 $_{-16.7}^{+16.4}$ &   2.2 $_{-28.1}^{+31.7}$ &     Halo \\
 & & & & & & & & & & & & \\ 
 7 & 0.57 $_{-0.05}^{+0.06}$ & 14.3   $_{-1.6}^{+1.8}$ &  4.0  $_{-0.2}^{+0.1}$ &  2.1  $_{-0.4}^{+0.7}$ &  489.9   $_{-109.4}^{+151.6}$ &  1451.1    $_{-42.1}^{+39.9}$ &   35.8    $_{-4.3}^{+9.8}$ &  -148391  $_{-5055}^{+4764}$ &  -60.9 $_{-8.8}^{+8.3}$ & -146.4 $_{-12.2}^{+10.6}$ &   6.0 $_{-12.8}^{+11.8}$ &     Disk \\
 & & & & & & & & & & & & \\ 
 8 & 0.52 $_{-0.24}^{+0.30}$ & 30.5   $_{-7.4}^{+13.4}$ &  9.6  $_{-6.9}^{+14.8}$ & 17.2  $_{-6.0}^{+6.5}$ &  882.5   $_{-462.6}^{+568.7}$ & -2744.0  $_{-3106.5}^{+2195.0}$ &  632.8  $_{-383.4}^{+453.9}$ &  -110197  $_{-18420} ^{+23747}$ &  109.1 $_{-49.6}^{+50.3}$ & -280.2 $_{-58.2}^{+56.9}$ &  -84.2 $_{-37.4}^{+33.3}$ &  Sequoia \\
 & & & & & & & & & & & & \\ 
 9 & 0.76 $_{-0.06}^{+0.06}$ & 30.7   $_{-4.4}^{+6.1}$ &  4.1  $_{-0.5}^{+0.4}$ &  4.0  $_{-1.3}^{+2.1}$ & 1649.8   $_{-404.9}^{+566.7}$ &  1830.3   $_{-125.6}^{+96.3}$ &   55.5   $_{-12.8}^{+17.5}$ &  -114866  $_{-6481}^{+7743}$ &  178.5 $_{-20.3}^{+22.7}$ &   99.1  $_{-8.6}^{+9.3}$ &   -2.5  $_{-0.7}^{+0.7}$ &     Halo \\
 & & & & & & & & & & & & \\ 
10 & 0.38 $_{-0.02}^{+0.03}$ &  9.1   $_{-0.1}^{+0.1}$ &  4.1  $_{-0.2}^{+0.2}$ &  2.4  $_{-0.2}^{+0.2}$ &  159.9    $_{-17.4}^{+20.5}$ &  1198.2    $_{-45.3}^{+38.8}$ &   82.4    $_{-9.5}^{+11.6}$ &  -166959   $_{-216}^{+229}$ &   34.7 $_{-0.7}^{+1.0}$ &  -52.8 $_{-2.3}^{+2.1}$ &    7.0  $_{-3.0}^{+2.9}$ &     Disk \\
            \hline
        \end{tabular}
        \end{adjustbox}
	    
	    \label{tab:orb}
	\end{table*}
	
\section{Discussion}\label{sec:discussion}

	In the following sub-sections we discuss in detail the results for the 475 very metal-poor stars analyzed. We will focus specifically on three key aspects. The first is the relation between the stars in our sample and the recently described remnants of the postulated \textit{Gaia Sequoia} and \textit{Gaia Sausage} accretion events \citep{belokurov2018,myeong2019}. For the sake of this analysis, following the hypothesis of \citet{belokurov2018} and \citet{myeong2019}, we assume that these accretion events are distinct, but see \citet{helmi2018} for an alternative view, particularly of \textit{Gaia Enceladus} as a single ancient major merger event.  The purpose of our work, however, is not to discern between the scenarios proposed to explain these structures in the Galactic halo, but rather to investigate their very low-metallicity content. %\citep[see, e.g.,][]{helmi2018,belokurov2018}. 
	The second key point is the analysis of low-metallicity stars with disk-like orbital properties that likely have a fundamental role in contributing to the understanding of the formation and evolution the MW's disk.  Finally, we discuss the properties and potential origin of the stars in our sample that are either loosely bound or not bound to the Galaxy.

	\subsection{\textit{Gaia Sausage} and \textit{Gaia Sequoia} candidate members}\label{sub:gsge}
	
		The exquisite data provided by Gaia DR2 has recently revealed the trace of at least two early major accretion events in the history of our Galaxy, referred to as \textit{Gaia Sausage} and \textit{Gaia Sequoia} \citep{helmi2018, belokurov2018, mackereth2019, myeong2019,koppelman2019}. These discoveries are a direct consequence of the development of computational techniques and resources capable of processing very large data sets.  %Specifically, action space, largely unexplored until relatively recently, is the ideal plane in which to use large samples of MW stars to identify and study possible sub-structures and debris from accretion events. The reason is that the actions are nearly conserved under the hypothesis that the potential is smoothly evolving \citep{binney1984}.

		Here we exploit the action-space classification provided in \citet[][their Figure~9]{myeong2019} to identify possible members of these accretion features within our sample of low-metallicity stars. 
		\citet{monty2019} have adopted a similar approach finding possible members of these systems with metallicities as low as $\rm [Fe/H]=-3.6$ dex. The number, abundances and abundance ratios of these stars could provide important information on the early evolution of the progenitors of the two accretion events.

\begin{figure*}
		 	\centering
		 	\includegraphics[width=16cm, trim={1cm 1.5cm 1cm 0cm}, clip]{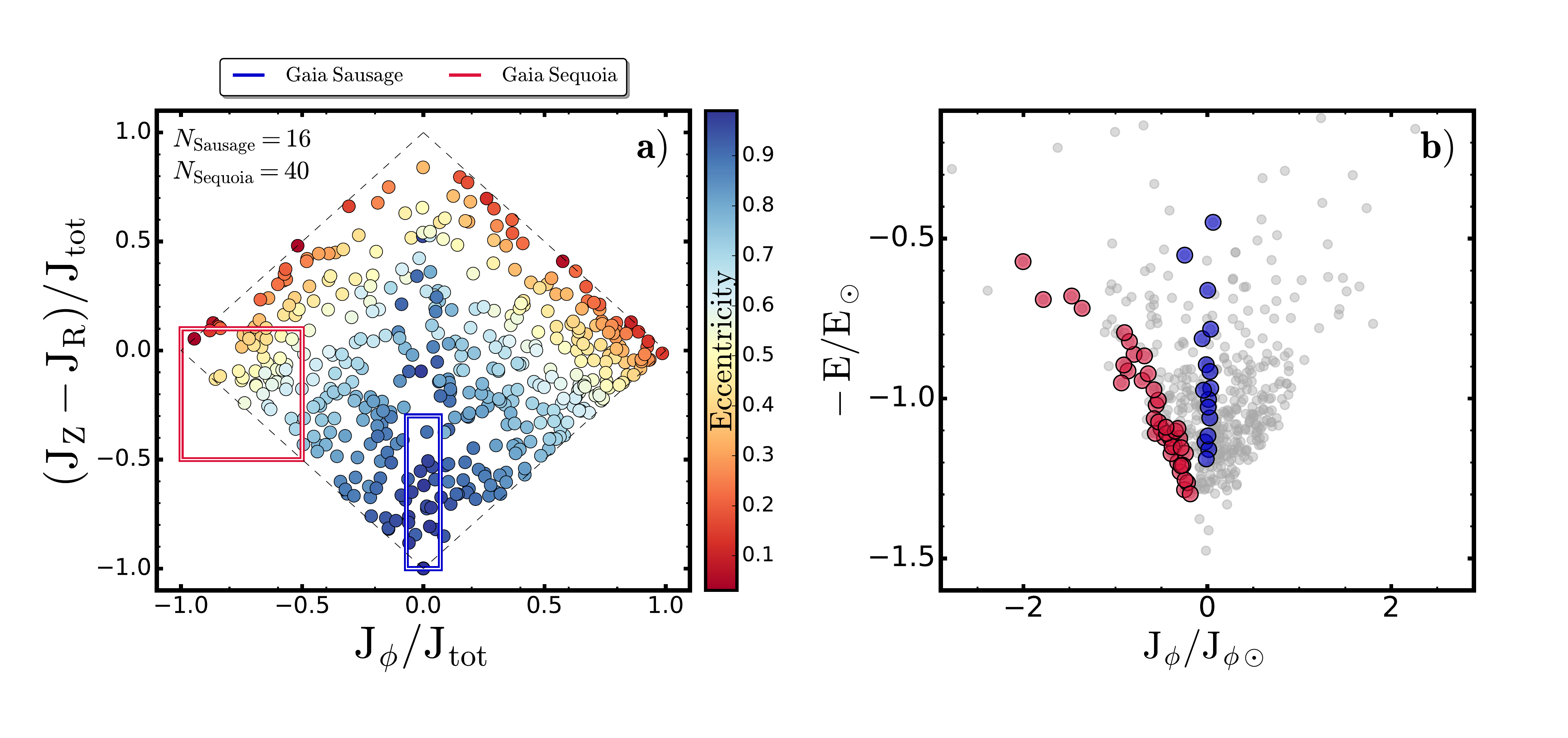}
		 	\includegraphics[width=16cm, trim={1cm 1.0cm 1cm 2cm}, clip]{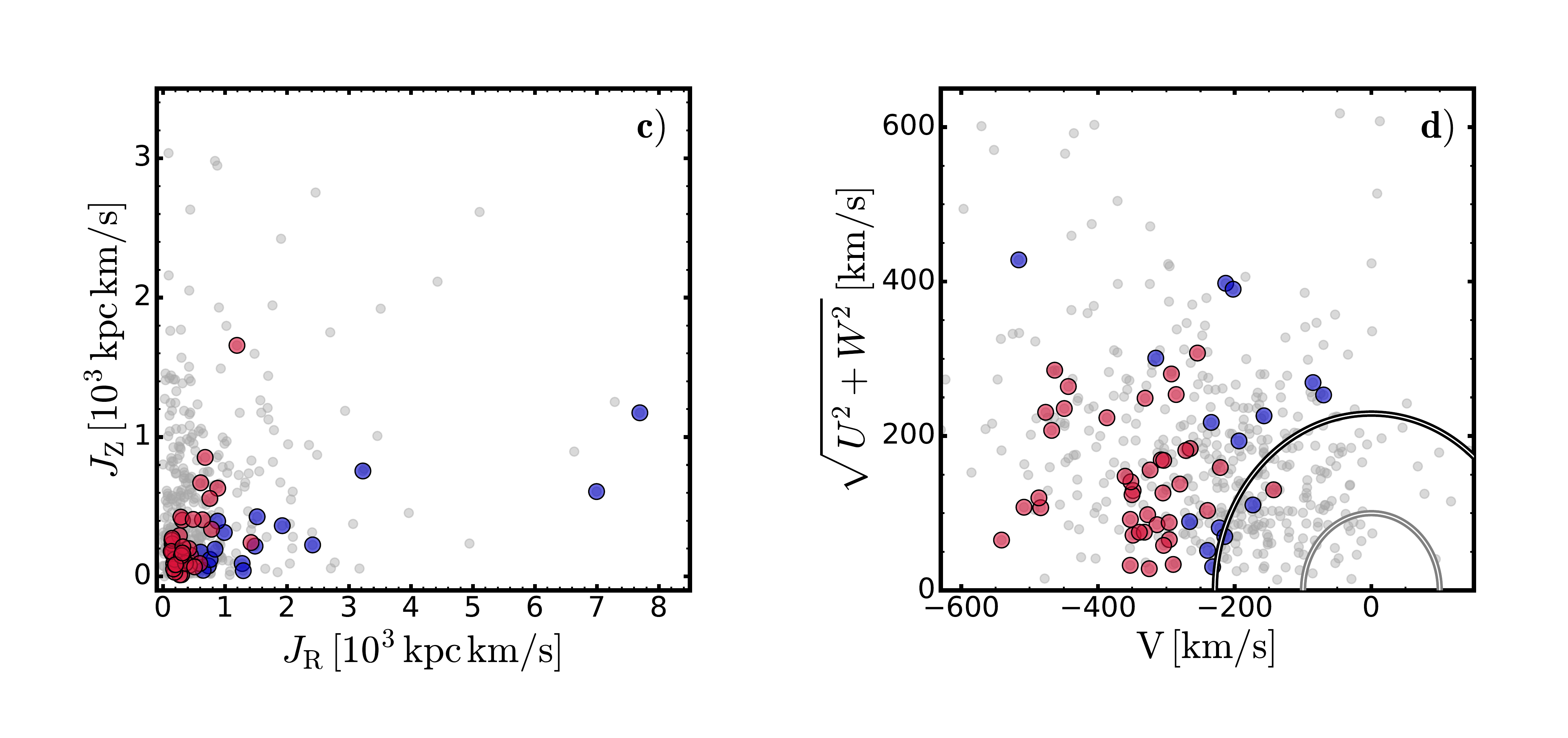}	\includegraphics[width=16cm, trim={1cm 1.0cm 1cm 2cm}, clip]{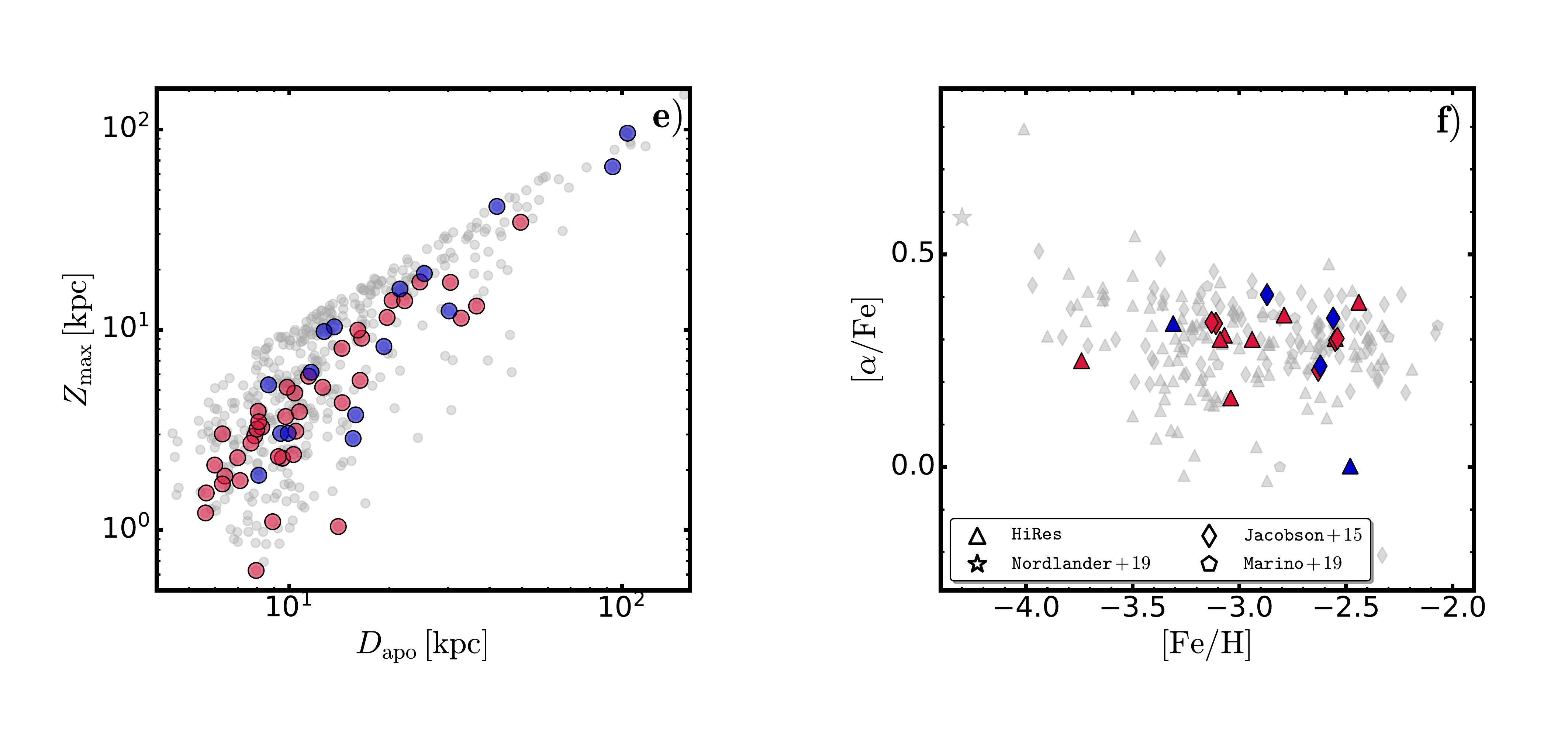}		 	
 	
		 	\caption{\textit{Panel a)} Action map for all the stars in our sample. The red and blue boxes identify the \textit{Gaia Sequoia} and \textit{Gaia Sausage} loci, as determined in \citet{myeong2019}. Each star is colour coded according to its eccentricity. 
		 	\textit{Panel b)} Energy $(E)$ against azimuthal action $(J_{\rm \phi})$ normalized by the solar values. Red and blue circles represent \textit{Sequoia} and \textit{Sausage} candidate members, respectively, while grey small points mark stars outside of the selection boxes in the action map. \textit{Panel c) and d)} Vertical action $(J_{\rm Z})$ against radial action $(J_{\rm R})$ and Toomre diagram, respectively.  The solid lines in panel d) show circular velocities of 100 and 239\,$\rm km\,s^{-1}$. \textit{Panel e)} Maximum altitude $(\rm Z_{\rm max})$ against apogalacticon distance $(D_{\rm apo})$. 
		 	\textit{Panel f)} Chemical abundances for all the stars in the \texttt{HiRes, Jacobson+15} and \texttt{Marino+19} samples, shown in grey shaded  triangles, diamonds and pentagons, respectively. The star SMSS~J160540.18--144323.1 \citep{nordlander2019}, shown with a star-like symbol, is arbitrarily put at [Fe/H]=--4.3 for plotting purposes as it is much more metal-poor than any of the other stars plotted. \textit{Gaia Sausage} and \textit{Gaia Sequoia} member candidates are marked with blue and red symbols, respectively. The values of $\rm [\alpha/Fe]$ have been computed as the mean of [Ca/Fe], [Mg/Fe], [TiI/Fe] and [TiII/Fe], whenever available.
		 	}
		 	\label{fig:gSgE}
		\end{figure*}

		The top-left panel of Figure~\ref{fig:gSgE} shows the action map $(J_{\rm Z}-J_{\rm R})/J_{\rm tot}$ vs. $J_{\rm \phi}/J_{\rm tot}$ with $J_{\rm tot}$ being the sum of the absolute value of the three actions $(J_{\rm tot} = J_{\rm R}+J_{\rm Z}+|J_{\rm \phi}|)$. Following the classification in \citet{myeong2019}, we highlight the loci of the \textit{Sequoia} and \textit{Sausage} accretion events with red and blue rectangles, respectively. 
		
		We find that out of the 475 analyzed stars, 16 stars are kinematically coincident with the \textit{Sausage} accretion event, while 40 stars are candidate \textit{Sequoia} members. 
		As expected from their definition and the action map \citep{helmi2018, belokurov2018, myeong2019, yuan2019}, the latter are characterized by mildly eccentric $(e\sim0.5)$ retrograde orbits, while the former have highly eccentric orbits $(e\sim 0.9)$. Appendix~\ref{app:orbits} shows some typical orbits for stars identified as possible \textit{Gaia Sequoia} and \textit{Gaia Sausage} members.
	
		We remind the reader that our membership identification follows the criteria introduced in \cite{myeong2019}, and is thus entirely based on the dynamics through the use of the action map. We stress that this approach does not allow for any ``background'' population that may be present in these regions of the action map.  Consequently, we cannot straightforwardly assume that all the stars in our sample that are dynamically coincident with the \textit{Sequoia}/\textit{Sausage} accretion events actually belong to such remnants. In Appendix~\ref{app:clustering} we have attempted to perform a more accurate analysis through the use of a clustering algorithm approach. Briefly, the clustering analysis of our very metal-poor sample does provide independent evidence for the existence of groupings consistent with the \textit{Sequoia} (group 6) and \textit{Sausage} (group 8) dynamical definitions, though there are also indications that our \textit{Sequoia} and \textit{Sausage} samples, as defined in Fig.\ \ref{fig:gSgE}, are potentially contaminated by a ``background'' population that might be as much as $\sim$~50\% and $\sim$35\%, respectively. These background estimates are determined by exploiting the clustering analysis groupings discussed in Appendix~\ref{app:clustering}, and the numbers of stars within the \textit{Gaia Sequoia} and \textit{Gaia Sausage} loci.% are populated by more than coherent group of stars, namely group \#5 and \#6 for the former, and groups \#3 and \#8 for the latter.} 

	Panels b), c), d) and e) of Figure~\ref{fig:gSgE} show a detailed analysis of stars identified as candidate \textit{Sequoia}, shown in red, and \textit{Sausage} members, shown in blue. 
	In panel c) we note that, by construction, \textit{Sausage} stars are characterized by more radial orbits, although at low 
	$J_{\rm R}$, some candidate \textit{Sequoia} stars seem to share the similar values of $J_{\rm R}$ as \textit{Sausage} stars. The Toomre diagram in panel d) shows that both groups are consistent with halo dynamics, and again we note that there is some degree of overlap between the two groups of stars. As regards panel b), which shows the energy versus azimuthal action,
	\textit{Sausage} candidates show the distinctive vertical distribution, indicative of almost null
	azimuthal angular momentum, while \textit{Sequoia} stars are clearly highly retrograde, as expected. Comparing panel b) with \citet[][their Figure~2]{koppelman2019} we note that our accreted candidates span a wider range in energy. However, we note that the definition of \textit{Sequoia} and \textit{Sausage} parameters differs from work to work. Indeed, \citet{yuan2019} identifies \textit{Sausage} members that lie well outside the selection box of \citet{myeong2019} and the energy range of \citet{koppelman2019}. For the sake of our analysis, we choose to be consistent with the \citet{myeong2019} classification, although we stress again that a number of the candidates  
	may not actually belong to the remnants of the accretion events. 
	
	As regards abundances, we find that the most metal-poor star in our sample that is a candidate member of \textit{Sequoia} (SMSS~J081112.13-054237.7) has a metallicity of $\rm [Fe/H]=-3.74$, while the most metal-poor \textit{Sausage} candidate (SMSS~J172604.29-590656.1) has $\rm[Fe/H]=-3.31$ dex.  Both stars come from the \texttt{HiRes} sample so that the abundance uncertainty is of order 0.1 (excluding any systematic uncertainties such as those arising from the neglect of 3D/NLTE effects).  These values are quite consistent with the results of \citet{monty2019}.  In that work, which uses dwarf stars, the lowest metallicity star plausibly associated with \textit{Sequoia}, G082--023, has $\rm [Fe/H]=-3.59 \pm 0.10$ while the most metal-poor star plausibly associated with \textit{Sausage}, G064--012, has $\rm [Fe/H]=-3.55 \pm 0.10$ \citep{monty2019}. 
		
	Finally, for the stars in the \texttt{HiRes, Jacobson+15} and \texttt{Marino+19} samples, we are able to investigate the chemical patterns of the likely accreted stars. Panel f) of Figure~\ref{fig:gSgE} shows [$\rm \alpha$/Fe] vs.\ [Fe/H] for the 218 stars for which [$\rm \alpha$/Fe] values are available.  Specifically, [$\rm \alpha$/Fe] is computed as the unweighted mean of [Mg/Fe], [Ca/Fe], [TiI/Fe] and [TiII/Fe] where available\footnote{We note that while detailed abundances, including those for neutron-capture elements, have been published for the \texttt{Jacobson+15} and \texttt{Marino+19} samples, this not the case for the \texttt{HiRes} sample (Yong et al.\ in preparation). Consequently, we refrain from investigating other element ratios.}.  The star SMSS~J160540.18--144323.1 \citep{nordlander2019} has been arbitrarily plotted at a metallicity of [Fe/H]=--4.3 since otherwise it would be the only star with [Fe/H]<--5 in the panel. A visual inspection reveals that, with the single exception of SMSS~J111201.72-221207.7,
	all the \textit{Sequoia} and \textit{Sausage} candidates are $\alpha$-enhanced, and no other trend is evident. Specifically, in our sample of very metal-poor \textit{Sequoia} and \textit{Sausage} candidates, we see no
	evidence for a ``knee'' in the ([$\rm \alpha$/Fe], [Fe/H]) relation.  The metallicity of the knee marks the abundance where [$\rm \alpha$/Fe] begins to decrease with increasing [Fe/H] as the nucleosynthetic contributions from SNe~Ia become increasingly important.  Our result, however, is not inconsistent with the results of \citet{matsuno2019} and \citet{monty2019} who find evidence of the presence of a knee at higher abundances than any of the \textit{Sequoia} and \textit{Sausage} candidates plotted in Fig.\ \ref{fig:gSgE}. For example, \citet{monty2019} indicate that the knee in \textit{Sequoia} is at [Fe/H] $\approx$ --2 while that for  \textit{Gaia Sausage} is at [Fe/H] $\approx$ --1.6, values significantly more metal-rich than any of the candidates in Fig.\ \ref{fig:gSgE}.
	
	\subsection{A very metal-weak component in the Thick Disk?}\label{sub:disk}
	
		It has recently been shown \citep{sestito2019a,sestito2020, dimatteo2019, venn2020} that, in contrast to the commonly accepted view, a significant fraction ($\sim 20\%$) of very low-metallicity stars resides in the MW disk rather than in the halo.
		Here we find a similar result: for the stars in our sample 102 out of 475 ($\sim 21\,\%$) exhibit disk-like dynamics in having orbits that are confined to within 3 kpc of the plane of the Milky Way.
		
		The straightforward conclusion would be to propose these stars as supporting the existence of 
		%a very metal-weak Thick Disk (vmwTD), 
		an extension to yet lower metallicities of the proposed Metal-Weak Thick-Disk \citep[e.g.,][]{chiba2000}. However, a detailed analysis is required to discern the origin of these stars. Specifically, it is of key importance to understand if they are indeed disk-like stars, or if they are, for example, halo stars whose orbital plane happens to lie in the MW disk.
		
		We therefore explore a number of orbital parameters to shed light on the nature of these stars. Specifically, \citet[][and references therein]{sales2009} have shown how the orbital eccentricity can be used to probe the formation scenario of the Milky Way thick disk. Furthermore, as mentioned in the previous sections, the actions, and in particular the azimuthal action, provide important clues for the origin of a star.  Here we couple these two orbital properties to disentangle the origin of the very low-metallicity stars residing in the MW disk. 
		
		As noted above, we classify as ``disk'' stars those with $|Z_{\rm max}|\leq 3\,\rm kpc$. This choice is %partially arbitrary and it is 
		made on the basis of the following considerations. First, \citet{li2017} find that the (exponential) scale height of the MW thick disk is $z_{\rm 0}=0.9\pm 0.1\,\rm kpc$; it then follows that the vast majority of the thick disk population should be found within $\sim$3 scale heights, i.e., within $|Z_{\rm max}|$ = 3 kpc.  
		
\begin{figure}
            \centering
	        \includegraphics[width=9cm, trim={1cm 4cm 1cm 14.55cm}, clip]{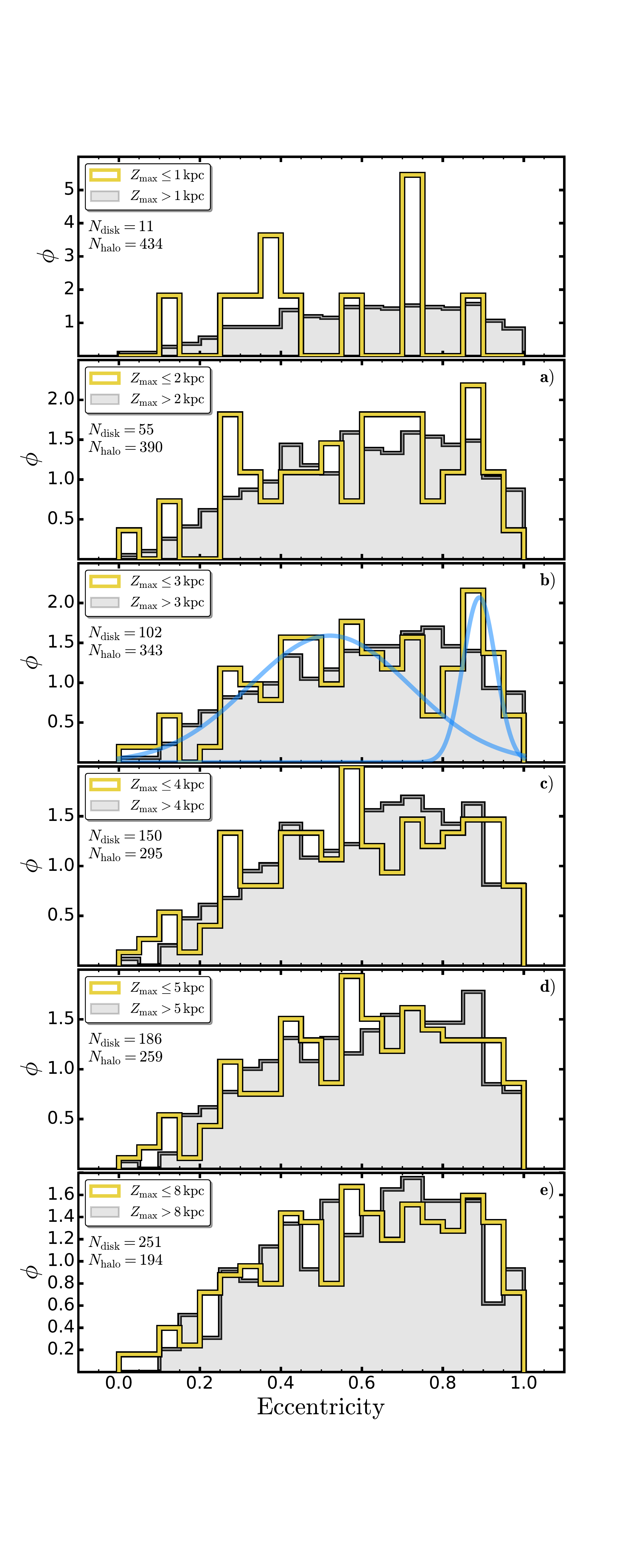}
	  	    \caption{Eccentricity distribution of disk-like stars and halo-like stars for different choice of the cutoff $|Z_{\rm max}|$, from $2\,\rm kpc$, panel a), to $8\,\rm kpc$, panel e). In each panel, the eccentricity distribution for stars within $|Z_{\rm max}|\leq N\,\rm kpc$ is shown with the yellow histogram, while stars with $|Z_{\rm max}| > N\,\rm kpc$ are indicated by the grey shaded histogram.  Stars with $D_{\rm apo} > $ 250~kpc are not considered.  Panel b) also shows, as blue continuous curves, the outcome of applying Gaussian mixture modeling to the set of $e$-values for the disk-like stars, i.e., without any binning. }
	  	    \label{fig:eccDist}
	    \end{figure}

		Second, Figure~\ref{fig:eccDist} shows the eccentricity distribution for different values of the maximum vertical excursion, from $2\,\rm kpc$ to $8\,\rm kpc$. For each panel, the yellow histogram (designated as ``disk'') shows the distribution of stars within $N\,\rm kpc$, with $N=\{2,3,4,5,8 \}$, while the grey shaded histogram (designated as ``halo'') represents stars with $|Z_{\rm max}|> N\,\rm kpc$. As is evident from the figure, for heights above the plane exceeding 3~kpc, i.e., panels c), d) and e), the eccentricity distributions for the stars above and below the cut-off height become increasingly similar. On the other hand, for a cutoff value of $|Z_{\rm max}|=3\,\rm kpc$, an apparent difference is present in the sense the $e$-distribution for the low $|Z_{\rm max}|$ stars has a possible excess of intermediate eccentricity stars together with a possible narrow surfeit of stars with $e$ $\approx$ 0.85, which is also evident in panel a).  However, application of both Kolmogorov-Smirnoff and  Anderson-Darling tests 
		\citep[see, e.g.][]{adtest} to compare the ``disk'' and ``halo'' distributions in panel b) revealed that the apparent differences are not statistically significant. Nonetheless, we adopt $|Z_{\rm max}| = 3\,\rm kpc$ as the value of $|Z_{\rm max}|$ to discriminate between predominantly disk and predominantly halo populations. For completeness, we also note that if we choose $|Z_{\rm max}|$ cutoff values of 2.5 or 4 $\rm kpc$ and repeat the analysis discussed below, the outcomes are essentially unaltered.
		
		%\GC{As we can see, for heights above the plane exceeding 3~kpc, i.e., panels d), e) and f), the eccentricity distributions of stars above and below the cut-off height become increasingly similar. On the other hand, when we adopt $|Z_{\rm max}|=3\,\rm kpc$ as the cutoff value (panel c) in Figure~\ref{fig:eccDist}, we find some small differences in the two distributions. However, both the Kolmogorov-Smirnoff and the Anderson-Darling determined such differences to be statistically not significant. Nonetheless,  we adopt $Z_{\rm max} = 3\,\rm kpc$ as the cutoff altitude to discriminate between disk and halo populations. For consistency, we repeated the same analysis for different choices of the cutoff radius, namely 2.5 and 4 $\rm kpc$, finding the same results. \\}
		Finally, the third reason for adopting a value of $|Z_{\rm max}| = 3\,\rm kpc$ for the disk-like population is that it is consistent with \citet{sestito2019a,sestito2020}, allowing our results to be directly compared with theirs.

		Looking again in detail at panel b) in Figure~\ref{fig:eccDist}, we can see that the eccentricity distribution of the disk-like stars hints at the presence of two main groups.  The first group has a relatively broad distribution peaking at $e \approx 0.55$ while the second population has a narrower distribution centred at $e \approx 0.85$.   This interpretation is confirmed by the application of Gaussian Mixture Modeling to the $e$-distribution for the stars with $|Z_{\rm max}|\leq 3\,\rm kpc$, a process that does not require any choice as regards histogram bin size. The best-fit is for two Gaussians, one centred at $e = 0.52$ containing 80\% of the population and with a standard deviation of 0.14. The second Gaussian is centred at $e = 0.89$ with a narrow $\sigma$ of 0.03. The two Gaussians are overplotted with blue thick lines in panel b) of Fig.~\ref{fig:eccDist}. We shall refer to these two groups as the ``low-eccentricity'' and  ``high-eccentricity'' populations, respectively, and
		adopt $e=0.75$ as the eccentricity to separate them. 
		
\begin{figure}
            \centering            
	        \includegraphics[width=8.5cm, trim={0cm 0cm 0cm 1cm}, clip]{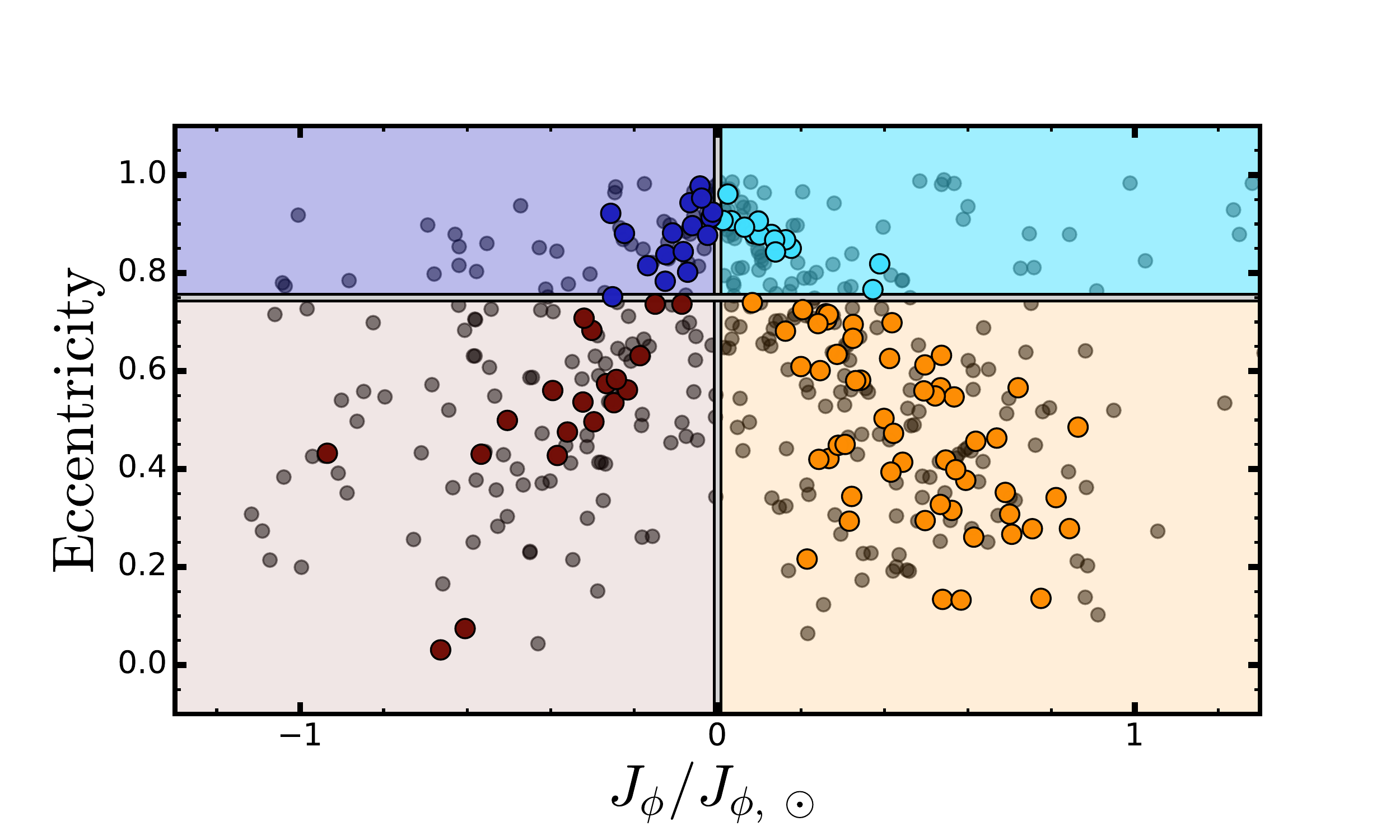}
	  	    \caption{Eccentricity vs. azimuthal actions for all the stars in our sample (grey dots). Coloured filled points are for stars with $Z_{\rm max} \leq 3\,\rm kpc$.  Specifically, stars on retrograde orbits with eccentricity greater and lower than $0.75$ are marked with dark-blue and dark-red circles, respectively. Prograde stars with eccentricity greater and lower than $0.75$ are indicated with azure and orange circles.}
	  	    \label{fig:EccJphi}
	    \end{figure}

		%As mentioned in previous sections, actions are an exquisite tracer of the origin of Milky Way stars.
		We now employ $J_{\rm \phi}$ to identify the motion of the disk stars as either prograde or retrograde. This is shown in Figure~\ref{fig:EccJphi}, where we mark retrograde high-$e$ and low-$e$ population stars with dark-blue and dark-red points, respectively, while prograde high-$e$ and low-$e$ group stars are indicated by azure and orange circles.\footnote{The colours are chosen consistently with the colour bar in Figure~\ref{fig:gSgE}, so that high eccentricity stars are identified by blue-ish colours, while red-ish colours indicate low eccentricity stars.} 
        Overall we find 72 low-$e$ stars (53 prograde and 19 retrograde) and 30 high-$e$ stars (15 prograde and 15 retrograde).
        
\begin{figure*}
            \centering            
	        \includegraphics[width=19cm, trim={1cm 0cm 0cm 0cm}, clip]{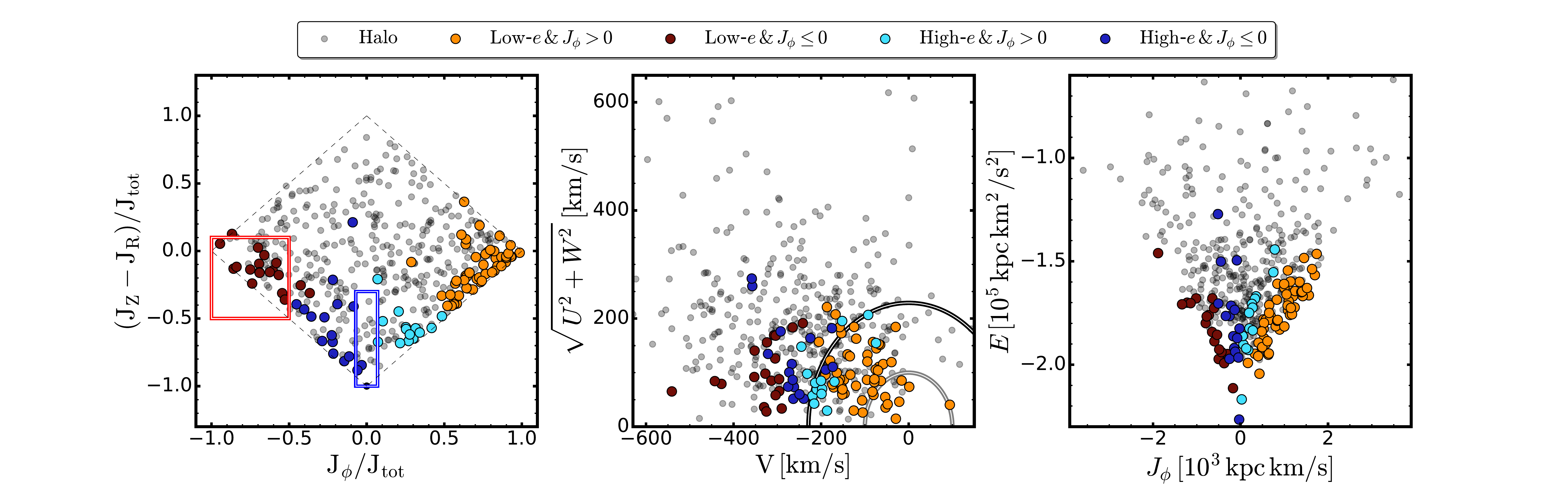}
	  	    \includegraphics[height=15cm, trim={0cm 0cm 3.5cm 0cm}, clip]{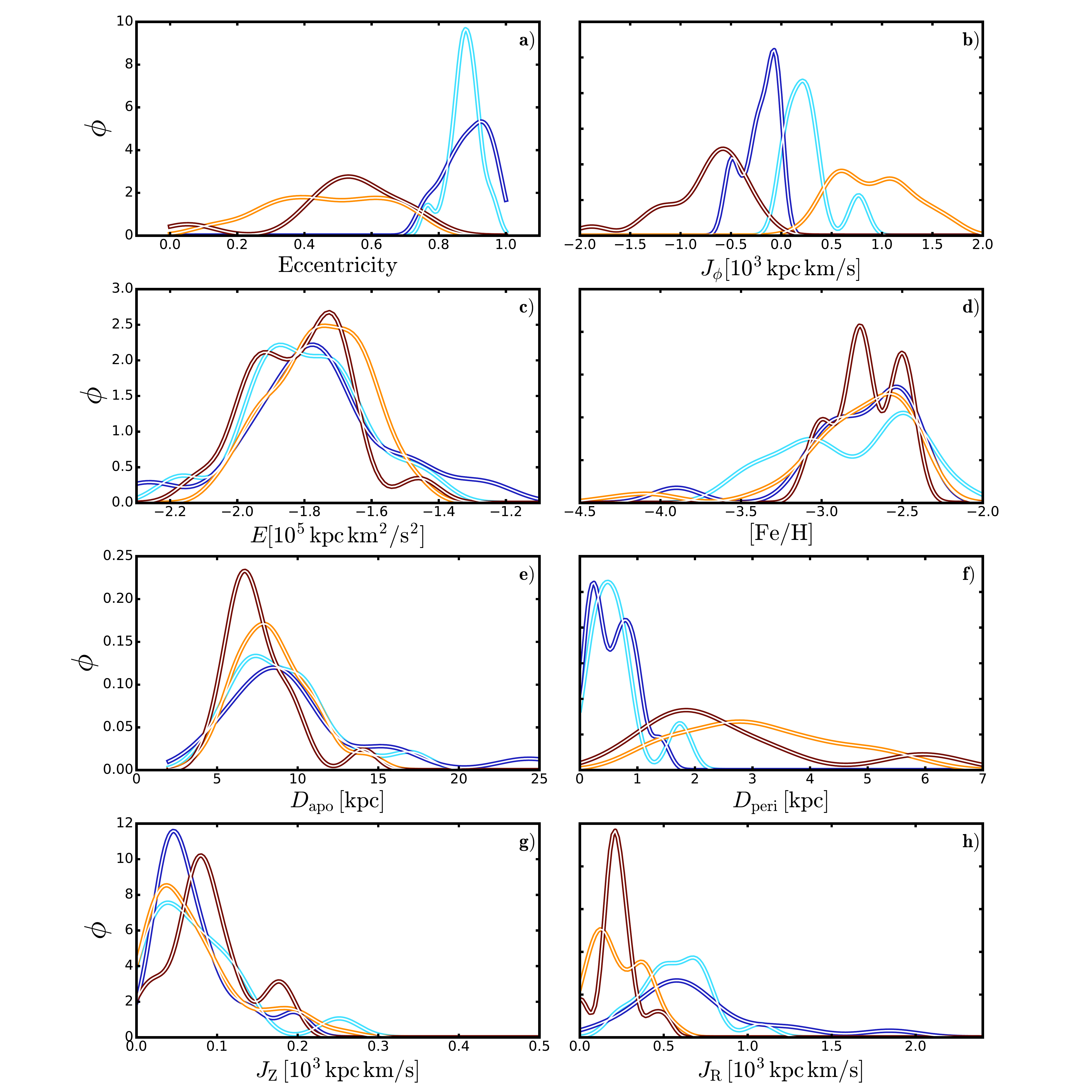}
	  	    \caption{\textit{Top panels.} Action map, Toomre diagram and energy vs. angular momentum for the stars with $|Z_{\rm max}| \leq 3\,\rm kpc$. The colours are the same as illustrated in Figure~\ref{fig:EccJphi}. \textit{Panels a) - h).} Kernel density function of the orbital parameters of the identified four groups of disk stars.}
	  	    \label{fig:DiskAnalaysis}
	    \end{figure*}

		We then analyzed the orbital parameters of each of these sub-groups of disk-star candidates. The results are shown in Figure~\ref{fig:DiskAnalaysis}. The distributions shown in panels a) -- j) are kernel density distributions computed by adopting a Gaussian kernel with a fixed value of 0.4 for the bandwidth scaling parameter, while 
		the top three panels show the action map, the Toomre diagram, and the $E$ vs. $J_{\rm \phi}$ plot,  respectively.  The disk candidates are marked with different colours, as defined in Figure~\ref{fig:EccJphi}. 
		
		\subsubsection{Low-eccentricity stars}

        In panel a) of Figure~\ref{fig:DiskAnalaysis}, it is interesting to see that the eccentricity distributions of the prograde and retrograde high-$e$ groups are almost identical, while, conversely, there are hints of a difference in the corresponding distributions for the low-$e$ stars.
		%The first interesting result is shown in panel a) of Figure~\ref{fig:DiskAnalaysis}. While the eccentricity distributions of the prograde and retrograde high-$e$ groups are almost identical, there are hints of a difference in distribution for the low-$e$ stars.
		Specifically, the prograde low-$e$ stars (orange line) have a broader distribution while the retrograde low-$e$ stars (red line) have a narrower distribution peaking at $e\sim0.5$-0.6.  Together these $e$-distributions are quite consistent with the theoretical results in \citet{sales2009}, particularly as regards the 
		$e$-distributions in the top-left panel of their Figure~3 \citep{sales2009}.  In that context the
		retrograde low-$e$ stars can be interpreted as an accreted population, while the prograde low-$e$ stars are likely ``in-situ'', i.e., born within the thick-disk of the Galaxy. 
		
		To support this interpretation we consider again the action map, here shown in the upper-left of Figure~\ref{fig:DiskAnalaysis} with the $|Z_{\rm max}| \leq 3\,\rm kpc$ stars identified.  It is evident from this panel that the majority of the retrograde low-$e$ stars fall within the locus defining the {\it Gaia Sequoia} accretion event, consistent with these stars having an accretion origin. Comparing the corresponding panels in Figures \ref{fig:gSgE} and \ref{fig:DiskAnalaysis} for the Toomre diagram and the Energy $(E)$ against azimuthal action $(J_{\rm \phi})$ diagram, respectively, confirms the connection.
		
		Regarding the prograde low-$e$ stars, a substantial number of these fall in the region of the Toomre diagram 
		usually restricted to disk stars; their rotation velocities lag that of the Sun by relatively small amounts, less than 100 $\rm km\,s^{-1}$ in some cases.  We therefore conclude that the 
		prograde low-$e$ stars define a {\it very metal-weak} component to the Galaxy's thick disk.  This conclusion is supported by the eccentricity distribution of the stars, which agrees well with the eccentricity distributions for (more metal-rich) thick-disk stars shown in \citet[][their Figure~12 and 14]{li2017}.
		
\begin{figure}
		    \centering
		    \includegraphics[width=9cm, trim={5cm 2cm 5cm 1cm}, clip]{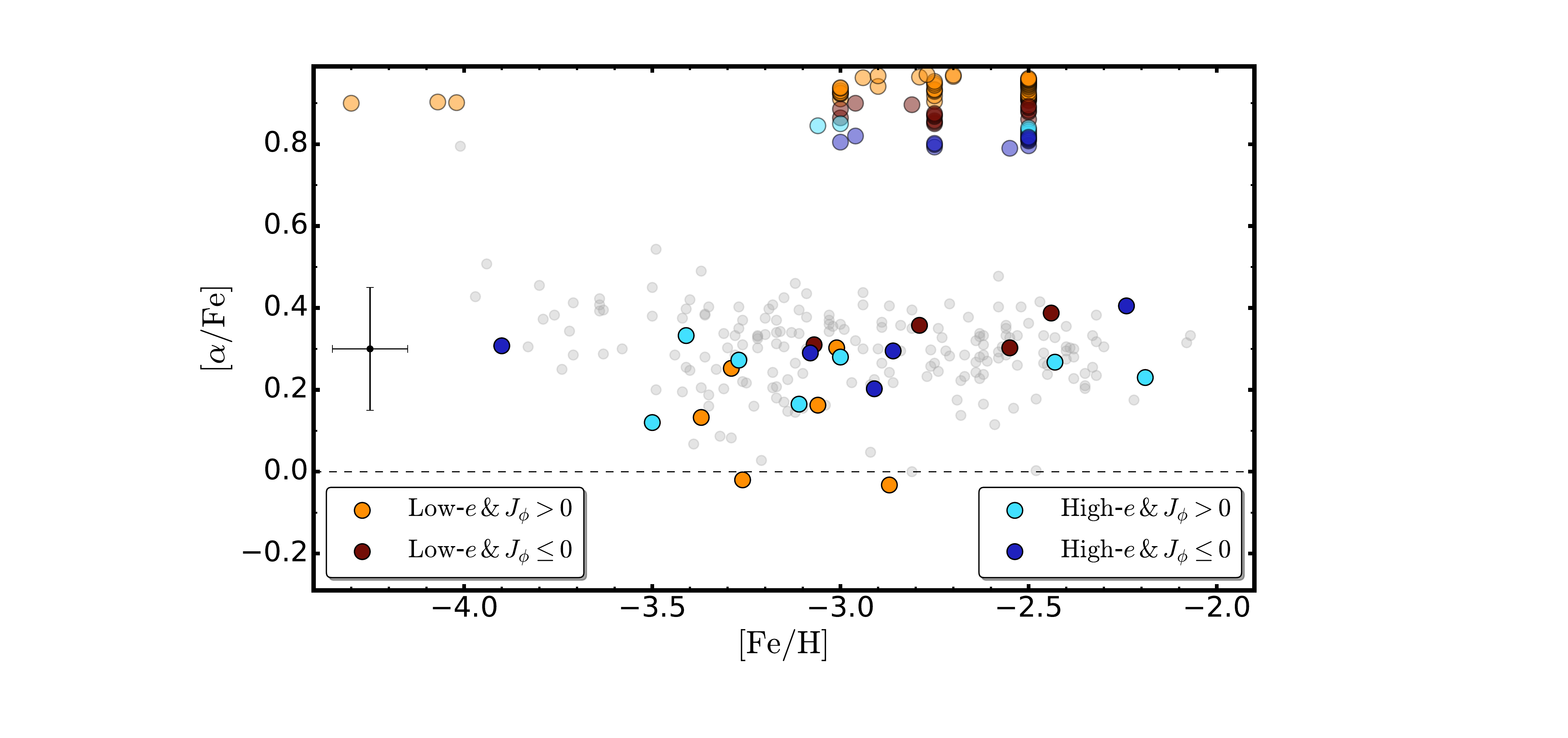}
		    \caption{[$\alpha$/Fe] vs. [Fe/H] for the stars in the \texttt{HiRes}, \texttt{Jacobson+15} and \texttt{Marino+19} samples. Low-$e$, low-$Z_{\rm max}$ prograde and retrograde stars are marked with orange and dark-red points, respectively, while high-$e$, low-$Z_{\rm max}$ prograde and retrograde stars are shown with azure and dark-blue colours.  Stars from the \texttt{LowRes} sample, for which only [Fe/H] values are available, are arbitrarily placed between [$\alpha$/Fe]=0.79 and [$\alpha$/Fe]=0.95 on the y-axis.  This is to allow an assessment of the [Fe/H] distributions of the samples. The (non-physical) range of [$\alpha$/Fe] values is needed to separate the points as the [Fe/H] values in the \texttt{LowRes} sample are generally quantized at 0.25 dex values.  %The actual [$\alpha$/Fe] values have no physical meaning. 
		    This also applies to the 3 orange points with [Fe/H] $\leq$ --4.0 that are from the \texttt{Sestito+19} dataset.  For the stars with high dispersion spectroscopic abundances, the typical uncertainty in [Fe/H] is $\pm$0.1 dex and $\pm$0.15 dex in [$\alpha$/Fe], as indicated by the black point on the left side.  For the stars from the {\texttt{LowRes} sample} the typical uncertainty in [Fe/H] is $\pm$0.3 dex.  The dashed line indicates [$\alpha$/Fe] = 0.} %\DM{Why?}\GC{Because for the LowRes sample the metallicity is quantized, so the points would just overlap.}}
		    \label{fig:lowprograde}
		\end{figure}

		These 53 low-$e$, low $Z_{\rm max}$ prograde stars represent $\sim$ 11\% of our total sample.  Of these 53, 6 are included in the high-dispersion data sets and the [$\alpha$/Fe] versus [Fe/H] for these stars is shown in Figure~\ref{fig:lowprograde}. Four of the 19 low-$e$, low $Z_{\rm max}$ retrograde stars are also included on the plot along with the remainder of the stars in the high-dispersion data sets.  For completeness as regards the [Fe/H] distributions of the samples, we also show in the upper part of the figure the [Fe/H] values for the remainder of the prograde and retrograde low-$e$, low $Z_{\rm max}$ samples.  Detailed abundance information, such as [$\alpha$/Fe], is not available for these stars that arise from the \texttt{LowRes} sample.  Further, in order to avoid any potential systematic effects, we have chosen not to plot the [$\alpha$/Fe] values for the 3 low-$e$, low $Z_{\rm max}$ \texttt{Sestito+19} sample stars in Figure \ref{fig:lowprograde}.  The stars are BD+44~493 ([Fe/H] = --4.30), 2MASS~J18082002-5104378 ([Fe/H] = --4.07) and LAMOST~J125346.09+075343.1 ([Fe/H] = --4.02) and all 3 have prograde orbits. The [Fe/H] values for these stars are taken from Table 1 of \citet{sestito2019a}.%\DM{Why plot the stars that have no available data?} \GC{We wanted to give an idea of the metallicity distribution of those stars}.
		
		It is evident from the figure that, though the sample is very limited, the four retrograde low-e stars show no obvious difference in location in this plane from the remainder of the (halo dominated) sample with high-dispersion abundance analyses.  The location of the six prograde low-e stars, however, is intriguing despite the small numbers.  It appears that the mean [$\alpha$/Fe] for these stars is lower than that for the full sample by perhaps 0.15, and two of the stars, namely SMSS~J230525.31-213807.0 which has [Fe/H] = \mbox{--3.26} and which is also known as HE~2302-2154a, and SMSS~J232121.57-160505.4 ([Fe/H] = --2.87, HE~2318-1621), are among the small number (of order a dozen in total) that have [$\alpha$/Fe] $<$ 0.1 in the high-dispersion samples.  Such $\alpha$-poor stars, also
		referred to as ``Fe-enhanced'' stars \citep[e.g.][]{yong2013,jacobson2015}, may reflect formation from gas enriched in SNe~Ia nucleosynthetic products that, if valid, may have implications for the epoch at which these stars settled into, or formed in, the thick disk \citep[see also][]{sestito2019a,sestito2020,dimatteo2019}. However, we consider further discussion of the element abundance distributions in these stars beyond the scope of the present paper.
		
		The overall [Fe/H] distribution of the retrograde and prograde stars as inferred from Figure \ref{fig:lowprograde} is similar to that for the full sample.  %We note, in particular, that the lowest abundance star in our prograde low-$e$, low $Z_{\rm max}$ group is the star BD+44~493 with [Fe/H] = --4.3, for which we find $e= 0.72$  and a $V$ velocity of --171.2 km/s, values that are quite comparable to those given in \citet{sestito2019a}.
		The star with the lowest combination of metallicity and eccentricity in our sample of prograde, low $Z_{\rm max}$ stars is SMSS~J190836.24--401623.5 that has [Fe/H] = --3.29 $\pm$ 0.10 and for which we find $e$ = 0.29 and a high $V$ velocity of --21 km~s$^{-1}$.  We also note that the orbital parameters derived here for the UMP star 2MASS~J1808002--5104378, which is included in our \texttt{Sestito+19} data set, are very similar to those found in \citet{sestito2019a}.  Specifically we find for this star $e$ = 0.13 and $V$ = --29 km~s$^{-1}$, while \citet{sestito2019a} list values of 0.09 and --45 km~s$^{-1}$, respectively.  Moreover, as noted by \citet{sestito2019a}, 
		the ``Caffau-star'' \citep{caffau2011}, which is an apparently carbon normal (i.e., [C/Fe] $<$ 0.7) dwarf (and therefore not included in our sample) with [Fe/H] $\approx$ \mbox{--5.0}, and which is the star with the lowest total metal abundance known to date, is a further example of an ultra-low metallicity star with a disk-like orbit. \citet{sestito2019a} determine that this star has a prograde orbit with $e$ = 0.12 that is confined to the Galactic plane, and which has a high $V$ velocity of --24 km~s$^{-1}$. We agree with the suggestion of  \citet{sestito2019a} that these stars may have formed in a gas-rich ``building-blocks'' of the proto-MW disk. The origin of these stars is also discussed within the context of the theoretical simulations presented in \citet{sestito2020b}. The simulations reveal the ubiquitous presence of populations of low-metallicity stars confined to the disk-plane.  In particular, the simulations show that the prograde planar population is accreted during the assembly phase of the disk, consistent with our interpretation and that of \citet{sestito2019a}.
		
		Panels b)-h) of Figure~\ref{fig:DiskAnalaysis} show the kernel density distributions of the other orbital parameters. These distributions do not exhibit clear differences between prograde and retrograde low-$e$ stars, with the possible exception of $D_{\rm apo}$, in panel e), and $D_{\rm peri}$, in panel f), which mimic the differences in the eccentricity distribution evident in panel a).
		
		We conclude that the low $Z_{\rm max}$ prograde and retrograde low-$e$ stars likely have different origins, with the former possibly being formed \textit{in-situ} in the Galaxy's thick disk while the latter are likely accreted from disrupted Milky Way satellites. Whether these latter stars belong to the \textit{Sequoia} main remnant, or its higher-energy tails \citep[Thamnos 1 and 2,][]{koppelman2019} is beyond the scope of the present work, although, as panel a) of Figure \ref{fig:DiskAnalaysis} shows, most fall within the region defining \textit{Sequoia} stars.

        \subsubsection{High-eccentricity stars}
        
		The interpretation of the 30 low $Z_{\rm max}$, high-$e$ stars in our sample is less straightforward, since, as the panels of Figure \ref{fig:DiskAnalaysis} reveal, nearly all their orbital parameters show similar distributions for the prograde and retrograde stars, with the only exception being the azimuthal action (by construction).  We note first that the only mechanism able to explain, at least qualitatively, the occurrence of disk stars with high-eccentricity orbits, is the heating mechanism discussed in \citet[][e.g., Figure 3]{sales2009}. Some of the stars in this sub-sample could therefore be disk stars heated by accretion events.  Alternatively, some could simply be halo stars that happen to have orbital planes that lie close to the Milky Way disk plane. 
		
		However, even though these high-$e$ stars do not satisfy the \textit{Gaia Sausage} \citet{myeong2019} membership criteria, we find that many qualitatively share its typical orbital properties: high-eccentricity, low or no angular momentum, small Galactic pericenters, Galactic apocenters as great as $\sim20$-$25\,\rm kpc$ and strong radial motions. We therefore argue that at least some stars could be associated with the \textit{Sausage} accretion event. In support of this conjecture, we note that they are consistent with the \citet{yuan2019} classification for \textit{Gaia Sausage} stars, both in the action map locus, and in the energy regime. 
		
		\citet{myeong2019} suggest that the \textit{Sausage} accretion event was an almost head-on collision with the Milky Way.  Such an event generates orbits with small perigalacticons that are strongly radial, eccentric, and with roughly equal numbers of prograde and retrograde stars.  These are the properties that we see for our low $Z_{\rm max}$, high-$e$ stars and it therefore seems reasonable to conclude that many of our low $Z_{\rm max}$, high-$e$ stars have their origin in the \textit{Sausage} accretion event.
		
		For seven of the 15 prograde high-$e$, low $Z_{\rm max}$ stars, detailed abundances are available from our high-dispersion data sets. The [$\alpha$/Fe] abundance ratios are shown as a function of [Fe/H] in Figure \ref{fig:lowprograde}.  The corresponding data for 5 of the 15 retrograde stars are also shown in the figure.  The [Fe/H] values for the remainder of the stars in these groups are shown across the top of the plot.  The numbers of stars with high-dispersion analyses in both high-$e$, low $Z_{\rm max}$ sub-samples are small, but there does not appear to be any obvious difference between them and the distribution of the full sample.

	    \subsection{The candidate unbound stars} \label{sub:escapers}

	    \begin{figure*}
            \centering            
	        \includegraphics[width=\textwidth, trim={0cm 0cm 0cm 0cm}, clip]{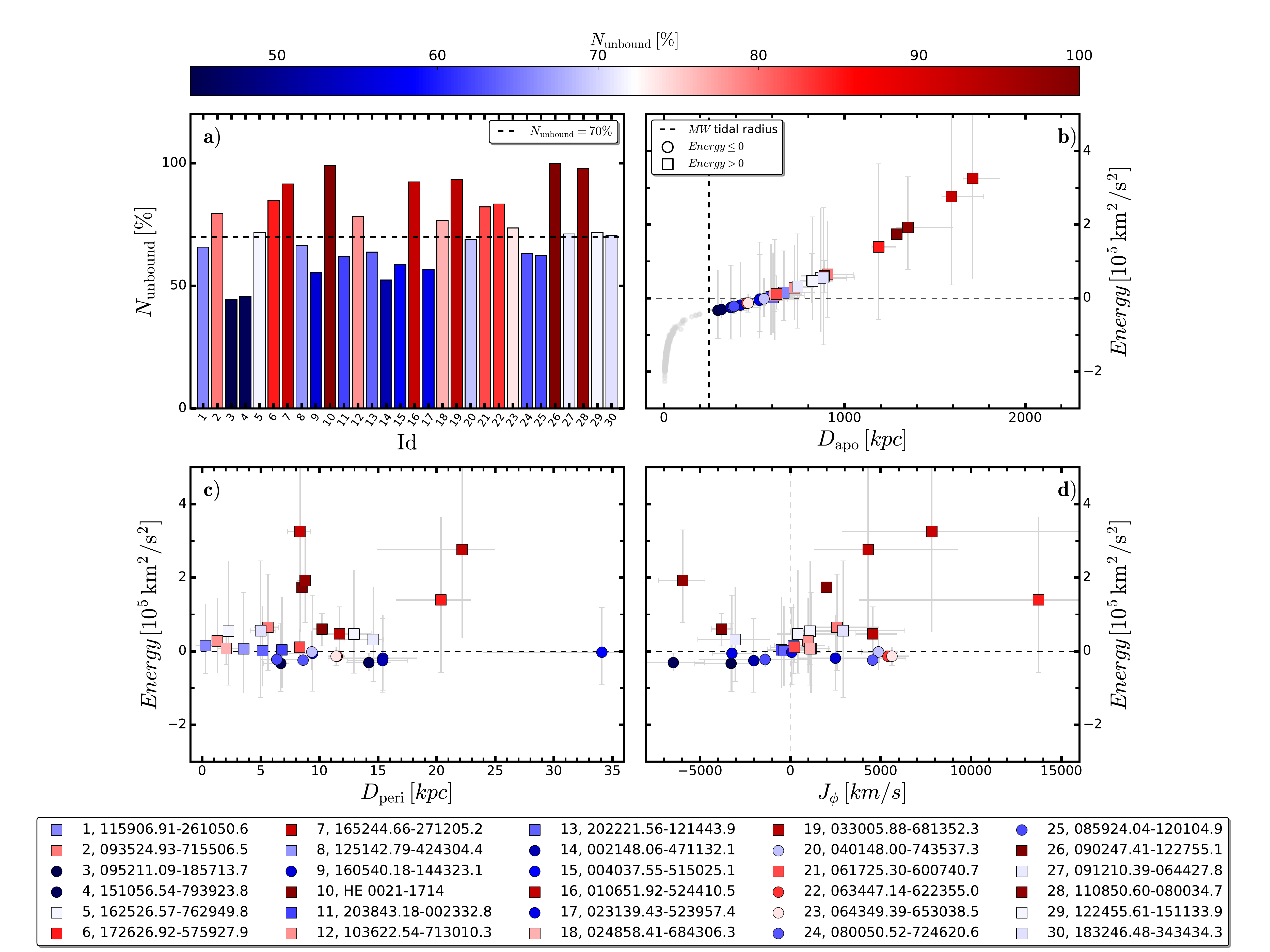}
	        \includegraphics[width=0.6\textwidth, trim={3cm 1cm 0cm 5.05cm}, clip]{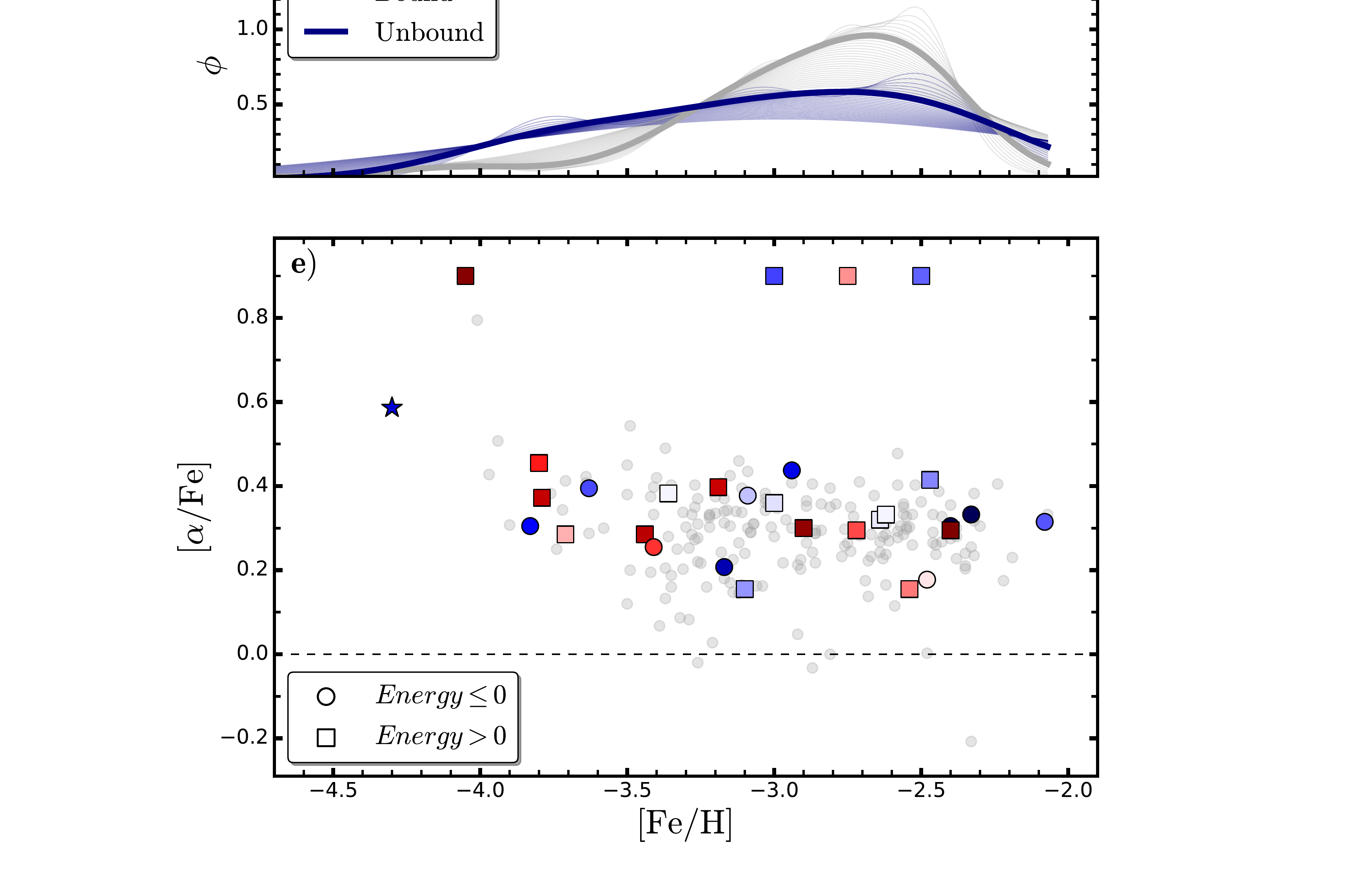}
	  	    \caption{\textit{Panel a).} Barplot of the percentage of unbound realizations for the 35 stars with apparent $D_{\rm apo} \geq 250\,{\rm kpc}$. The names of the stars are listed in the panel below panels c) and d). Each bar is colour-coded according to the percentage of $N_{\rm unbound}$, as shown in the top colour bar. \textit{Panel b).} Energy against apparent apogalacticon distance. Filled circles indicate stars with negative energy, while filled squares mark those with positive energy. The black vertical dashed line marks the Milky Way tidal radius (250\, kpc), while the stars with $D_{\rm apo}<250\,\rm kpc$ are shown as grey shaded points. \textit{Panel c)} and \textit{Panel d)} show the Energy against the perigalacticon distance $(D_{\rm peri})$ and the Energy against the azimuthal action $(J_{\rm \phi})$. Individual error bars are shown in panels b), c) and d). \textit{Panel e)}. [$\alpha$/Fe] vs. [Fe/H] for the candidate unbound stars that occur in the 
	  	    \texttt{HiRes, Jacobson+15} and \texttt{Marino+19} samples.  As in Fig.~\ref{fig:gSgE}, the star SMSS~J160540.18--144323.2 \citep{nordlander2019} is shown with a star-like symbol placed at [Fe/H]=--4.3. As for the other panels, filled circles indicate stars with negative energy, while filled squares mark those with positive energy.  The remainder of the stars in our high-dispersion samples are shown as light-grey circles.  The metallicity estimates for the 4 stars not in our high dispersion samples are plotted at the top of the panel. 
	  	    %\GDC{This figure may end up basically the same.  As far as I can see only SMSS J183246.48-343434.3 is in the questionable data set but it has a Gaia radial velocity so remains in the sample.  Only question will be if we still find it to be an escaper}.
	  	    }
	  	\label{fig:escapers}
	    \end{figure*} 

        As introduced in Section~\ref{sec:results}, we find 30 stars with potentially unbound orbits, defined by having an apparent $D_{\rm apo} \geq 250\,{\rm kpc}$\footnote{For a genuine unbound orbit $D_{\rm apo} = \infty$. The term ``apparent $D_{\rm apo}$'' employed here for the potentially unbound stars represents the Galactocentric radius at either the --2 Gyr or +2 Gyr endpoint of the orbit integration, whichever is larger.}. One star, HE~0020--1741, comes from our subset of the \texttt{Sestito+19} sample. \citet{sestito2019a} list $D_{\rm apo} \approx 296\,\rm kpc$ for this star, the largest value in their determinations, while in the \citet{mackereth2018} catalogue\footnote{\url{https://vizier.u-strasbg.fr/viz-bin/VizieR?-source=I/348}} the star is classified as unbound in accord with our result. The remaining 29 stars are from the SkyMapper samples.  
        
        To test the sensitivity of the results to the adopted potential we investigated the orbits of these stars using a different choice of the potential, namely the \texttt{GALPY} \textit{MWPotential2014}. The calculations reveal that all 30 stars again have $D_{\rm apo} > 250\,\rm kpc$. In addition, for this choice of potential, we find that the star SMSS~J044419.01--111851.2 may also be unbound; it is on a loosely bound orbit in the \texttt{McMillan2017} potential with an apogalacticon distance of \mbox{$\sim 200\,\rm kpc$}. 
        
        In order to shed light on the nature of these stars we have investigated the distributions of the 500 random realizations of the orbits, together with the corresponding orbital parameters, such as the apparent apo/peri-galacticon distance ratio $(D_{\rm apo}/D_{\rm peri})$, the energy ($E$) and the azimuthal action $(J_{\rm\phi})$. Panels a) -- d) of Figure~\ref{fig:escapers} show the information for the 30 potentially unbound stars. 
        
        Since the direct orbit integration for the observed positions and velocities resulted in unbound orbits (apparent $D_{\rm apo}>250\,\rm kpc$), it would be incorrect to adopt the uncertainties as defined in Section~\ref{sub:orb} for all parameters.  Specifically, for these stars we opt not to give uncertainties for the apparent $D_{\rm apo},\, e,\, Z_{\rm max}$ and $J_{\rm R}$ values since the medians for the 500 realizations and the ``observed'' values, i.e., the values from the observed properties, differ significantly. On the other hand, for the remaining orbital parameters, namely $D_{\rm peri},\, E,\, J_{\rm \phi},\, J_{\rm Z},$ and the  $U,\, V,\, W$ velocities, the median values are consistent with the observed ones, and therefore we compute the uncertainties in these quantities as before, from the difference between the $16^{\rm th}$ and the $84^{\rm th}$ percentile of the PDFs.     
        
        Panel a) of Figure~\ref{fig:escapers} shows the fraction of unbound realizations for each star. Each bar is colour-coded according to its percentage, from a minimum of 42.6\% for SMSS~J095211.09--185713.7 (number 3 in the identification panel in the Figure \ref{fig:escapers})  to a maximum of 100\% for SMSS~J090247.41--122755.1 (\#26). White- and reddish colour tones indicate stars with a fraction of unbound orbits greater than 70\%, which we take as a conservative value to identify likely unbound stars.
        Blueish colours represent stars with a lower unbound fraction. A visual inspection of panel b) reveals that nearly all stars with $N_{\rm unbound}>70\%$ exhibit positive energies, thus confirming that they are likely to be escaping from the Galaxy. Overall, we find that 17 stars have $N_{\rm unbound}>70\%$, and of these 15 have $E>0$\@ (the two stars with $N_{\rm unbound}>70\%$ but $E \leq 0$\@ are numbers 22 and 23).  Panel b) also shows that four stars (numbers 1, 8, 11, 13) have a positive energy, but with $N_{\rm unbound}$ slightly below 70\%. 
        We consider these stars as also likely unbound, bringing the total number of candidate unbound stars to 21, or 4.4\% of the total sample. We note in particular that aside from HE~0020--1741 (\#10), three further stars in our set of 21 unbound candidates are also classified as unbound in the \citet{mackereth2018} catalogue.  These are the bright $r$-process element enhanced star SMSS~J203843.18--002332.8 \citep[\#11, RAVE J203843.2--002333,][]{placco2017}, together with SMSS~J183246.48--343434.3 (\#30) and SMSS~J202221.56--121443.9 (\#13). On the other hand, one of our stars, SMSS~J103622.54-713010.3 (\#12), has a bound orbit in the \citet{mackereth2018} catalogue. There are no stars in common with the list of 20 `clean' high-velocity star candidates with unbound probability exceeding 70\% in \citet{marchetti18}.
        
        The remaining nine stars have negative (bound) energies, although the values are consistent with zero within their uncertainties. Their classification is thus uncertain as they could be unbound or on loosely bound orbits.  None are found in the \citet{mackereth2018} catalogue.
        
        We now speculate as to the origin of these stars, proposing three possible physical mechanisms that could provide each star with sufficient energy to escape the Galaxy. 
        \begin{itemize}
            \item A star in a close binary can be expelled from the Galactic Centre via an interaction with the central black hole.  The clearest example of this process is the star S5-HVS1 discussed in \citet{koposov20}. %\DM{e.g., Koposov et al. 2020?};
            \item A star can acquire high velocity (of order of the binary's pre-SNe explosion orbital velocity) from being in a binary when the companion explodes as a supernova \citep[e.g.,][]{eldridge11}.  
            \item A star can acquire energy as part of a gravitational interaction involving the merger of a dwarf galaxy with the Milky Way \citep[e.g.,][]{abadi09}.
        \end{itemize} 
        
        For the first mechanism to happen, the star has to have an origin close to the Galactic Centre, which means that its $D_{\rm peri}$ should be near zero.  However, as panel c) of Figure~\ref{fig:escapers} shows, only one star (SMSS~J115906.91--261050.6,
        \#1, $D_{\rm peri}$ = 0.26 kpc) has a perigalacticon distance within 1~kpc of the Galactic Centre, while the remainder  of the candidates have $D_{\rm peri} > 1\,{\rm kpc}$.  This suggests that the first possibility is unlikely, particularly when it is recognised that all the stars in the sample are giants and therefore unlikely to be in a sufficiently compact binary.
        
        The second mechanism also seems unlikely because the unbound stars are all giants with, as a consequence, relatively large stellar radii.  As a result, the separation between the components of any pre-SNe binary containing the star is unlikely to be sufficiently small that the orbital velocity, which underlies the ``kick velocity'' provided when the companion becomes a SNe, would be sufficiently high that the liberated star is no longer bound to the Galaxy.  %,and likely old, in which case it is hard to imagine what the binary companion star, which explodes as a supernova to provide the energetic ``kick'' could be.  Perhaps a SNeIa from a white dwarf companion but, 
        Furthermore, at least for those stars where high dispersion spectra are available, there is no evidence of any ``pollution'' from the SNe event. %\DM{What is the timescale for escape? i.e., how long did it take these stars to get to their present positions?  This gives an approximate time-scale over which all the SN explosions must have happened; I suspect the inferred rate would be much too high}
        
        This leaves us with the third possible origin, which is plausible given that it is generally accepted that the formation of the Galactic halo is driven by the accretion and tidal disruption of dwarf galaxies \citep[e.g., the recent discovery of remnants of accretion events:][]{helmi2018, belokurov2018, koppelman2019, myeong2019}. Specifically, we postulate that our set of unbound stars originated in the outskirts of dwarf galaxies that were accreted by the Milky Way, gaining energy from the gravitational interaction that resulted in the disruption of the dwarfs.  Given the relatively low metallicities of the unbound stars, which range from --4 (or less) to --2 in [Fe/H] (see the lowermost panel of Fig.\ \ref{fig:escapers}), we speculate that the disrupted systems were relatively low-mass, low-metallicity systems.In this context, we note that 16 out of 21  stars have prograde orbits, while 5  have negative $J_{\rm \phi}$ and thus a retrograde orbit.  This likely indicates that multiple accretion events may be involved. However, it is necessary to keep in mind that the timescale for an unbound star to reach the virial radius from the inner regions of the Galaxy is $\sim$ 1 Gyr.  Consequently, the gravitational interactions that generated the unbound stars in our sample likely occurred relatively recently, which may argue against the proposed ``origin in accretion events'' scenario.  Detailed evaluation of the orbits of the unbound stars, individually and collectively, is required to assess the situation and to investigate their origins(s).  The metallicities of other candidate unbound stars, such as those in \citet{marchetti18} will also provide important input \citep[e.g.,][]{hawkins18}.
        
        We note also that there is a fourth possibility, that uncertainties in the analysis lead to incorrect orbital parameters. For example, if the distance to the star used in calculating the orbit were overestimated, this could result in unbound or nearly-bound status. This is the likely explanation for the discrepancy concerning the star SMSS~J103622.54--713010.3 (unbound here, bound in the \citet{mackereth2018} catalogue) as our adopted distance is more than a factor of two larger than the \citet{bailerjones2018} distance.  The availability of improved parallaxes and stellar parameters from the forthcoming Gaia EDR3 and DR3 releases will help alleviate these discrepancies. As another possibility, we note that the total mass of the Galaxy may in fact be larger than that used in our modeling. If this is the case then, although on high-energy orbits, the stars would remain bound \citep[e.g.,][]{monari18,fritz20}.
        %It is also possible that the total mass of the Galaxy is larger than that used in our modeling, so that while on high-energy orbits, the stars remain bound \citep[e.g.,][]{monari18,fritz20}.
        
        Panel e) of Figure \ref{fig:escapers} shows the [$\alpha$/Fe] vs. [Fe/H] relation for 26 of the 30 candidate unbound stars that are in our high-dispersion samples, together with the values for the remainder of the stars from the high dispersion samples.  The estimates of [Fe/H] for the remaining 4 stars, from the \texttt{LowRes} data set, are shown at the top of the panel.  It is evident from this panel that candidate unbound stars are not distinguished from the full sample as regards the overall metallicity distribution or the [$\alpha$/Fe] distribution.

	\section{Summary and conclusions} \label{sec:conc}
		In this work we have analyzed a sample of 475 very metal-poor giant stars, most of which have originated from the SkyMapper search for the most metal-poor stars in our Galaxy \citep{dacosta2019}. The data set covers a  metallicity range of almost five dex $(-6.5 < {\rm [Fe/H]} \leq -2)$, and together with the relatively large number of stars, makes it ideal to investigate the kinematics, and ultimately the origin, of these very rare and important objects together with the implications for the formation of the Milky Way.
		
		We first exploited the action map for our sample together with the classification criteria of \citet{myeong2019} to identify candidate members of the \textit{Gaia Sausage} and \textit{Gaia Sequoia} accretion events.  We find 16 stars dynamically consistent with \textit{Gaia Sausage} and 40 with \textit{Gaia Sequoia}.  While we cannot be certain all candidates are in fact associated with these entities, the lowest metallicities ([Fe/H] = --3.31 for \textit{Gaia Sausage} and --3.74 for \textit{Gaia Sequoia}) are quite consistent with the findings of \citet{monty2019}. With a single exception, all our candidate \textit{Gaia Sausage} and \textit{Gaia Sequoia} stars for which we have high-dispersion spectra are $\alpha$-rich, similar to the general halo population.  This is again consistent with the results of \citet{monty2019}.
		
		The recent work of \citet{sestito2019a, sestito2020}, \citet{dimatteo2019} and \citet{venn2020} has revealed an unexpected significant population of very low-metallicity stars residing in the plane of the Galaxy. 
		We find a similar result in that $\sim21\,\%$ of the stars in our sample have orbits that remain confined to within 3\,$\rm kpc$ of the Galactic plane. Moreover, these stars show a different eccentricity distribution compared to the stars with larger $|Z_{\rm max}|$ values, pointing towards a different origin and/or evolution compared to the (halo dominated) bulk of the sample. 
		
		Our detailed analysis of these low $|Z_{\rm max}|$ stars reveals four sub-populations as regards orbit eccentricity and prograde or retrograde motion.  Of particular interest are the stars with relatively low eccentricities ($e < 0.75, {\rm median} \approx 0.5)$ and prograde velocities.  These stars, which make up $\sim$ 11\% of the total sample, have metallicities at least as low as [Fe/H] = --4.3 and are best interpreted as revealing the existence of a very low-metallicity tail to the Galaxy's metal-weak thick disk population \citep[e.g.][]{chiba2000}.  On the other hand, the low-$e$ retrograde stars that have $|Z_{\rm max}| \leq 3\,\rm kpc$ ($\sim$ 4\% of the sample) are most likely an accreted population.  We also find a population ($\sim$~6\% of the sample) of low 
		$|Z_{\rm max}|$ stars that have high eccentricity orbits (${\rm median} \approx 0.88$) with small pericenters and which are split equally between prograde and retrograde motion.  It seems likely that many of these stars might be associated with the 
		\textit{Gaia Sausage} accretion event \citep{myeong2019,yuan2019,koppelman2019}.
		With the possible exception of the low-$e$, low $|Z_{\rm max}|$ prograde stars that may have a somewhat lower mean [$\alpha$/Fe] abundance ratio, none of the four sub-populations with low $|Z_{\rm max}|$ are distinguished, as regards [$\alpha$/Fe] or [Fe/H], from the full set of stars for which high-dispersion based analyses are available. 
		
		Finally, we find that a small fraction of our sample (21 stars, $\sim$ 4.4\%) are likely to be escaping from the Galaxy, i.e., are on orbits that are not bound.  The [Fe/H] and [$\alpha$/Fe] distributions of these stars are not distinguished from those for the full sample; for example, their metallicities are spread from --4 (or less) to --2 in [Fe/H].  Our preferred interpretation for these stars is that they have acquired sufficient energy to escape from the Galaxy via the gravitational interaction that occurs when infalling dwarf galaxies are tidally disrupted by the Milky Way.

	\section*{Acknowledgements}

    We thank the referee for their comments on the original version of the manuscript. This work has received funding from the European Research Council (ERC) under the European Union's Horizon 2020 research innovation programme (Grant Agreement ERC-StG 2016, No 716082 `GALFOR', PI: Milone, \url{http://progetti.dfa.unipd.it/GALFOR}). A.D.M. is supported by an Australian Research Council Future Fellowship (FT160100206). A.F.M. acknowledges support by the European Union's Horizon 2020 research and innovation programme under the Marie Sk{\l}odowska-Curie (Grant Agreement No 797100).  A.R.C. is supported in part by the Australian Research Council through a Discovery Early Career Researcher Award (DE190100656). A.M.A. acknowledges support from the Swedish Research Council (VR 2016-03765), and the project grant “The New Milky Way” (KAW 2013.0052) from the Knut and Alice Wallenberg Foundation.  Parts of this research were supported by the Australian Research Council Centre of Excellence for All Sky Astrophysics in 3 Dimensions (ASTRO 3D), through project number CE170100013. A.P.M. acknowledges support from MIUR through the FARE project R164RM93XW SEMPLICE (PI: Milone) and the PRIN program 2017Z2HSMF (PI: Bedin). K.L. acknowledges funds from the European Research Council (ERC) under the European Union’s Horizon 2020 research and innovation programme (Grant agreement No. 852977). AF acknowledges support from NSF grant AST-1255160.
    %{\color{red} Co-authors: please add your grant support info}.\\
    
    This work has made use of data from the European Space Agency (ESA) mission \textit{Gaia} (https://www.cosmos.esa.int/gaia), processed by the \textit{Gaia} Data Processing and Analysis Consortium (DPAC, http://www.cosmos.esa.int/web/gaia/dpac/consortium).  Funding for the DPAC has been provided by national institutions, in particular the institutions participating in \textit{Gaia} Multilateral Agreement.

	\section*{Data availability}
	The data underlying this article are available in the article and in its online supplementary material.

%%%%%%%%%%%%%%%%%%%%%%%%%%%%%%%%%%%%%%%%%%%%%%%%%%



\bibliographystyle{mnras}
\bibliography{bibliography} % if your bibtex file is called example.bib

%%%%%%%%%%%%%%%%% APPENDICES %%%%%%%%%%%%%%%%%%%%%

\appendix

\section{Clustering analysis}\label{app:clustering}	
	As an alternative approach, we analyzed our sample of stars exploiting the \texttt{scikit-learn} Spectral clustering algorithm \citep{pedregosa2011}. 
	This analysis has the advantage of being almost completely independent from our choices (i.e., limiting $|Z_{\rm max}|$, prograde/retrogade etc.), while on the other hand has the limitations of being a ``blind'' analysis. Specifically, in order for the result to be trustworthy, we should have  knowledge of how the selection biases (e.g., bright stars in the solar neighbourhood, uncrowded stellar environments and the other effects discussed in \citet{dacosta2019}) propagate into the cluster choices. Unfortunately, it is not possible to quantify these biases.
	
	Nonetheless, we believe that it is worth exploring this fully independent classification approach. The clustering has been performed in the 4-dimensional space with the three actions and the eccentricity ($J_{\rm R},\, J_{\rm \phi},\, J_{\rm Z},\, e$) with the following input parameters: \texttt{affinity}$=$\texttt{nearest\_neighbors} and \texttt{assign\_labels}$=$\texttt{discretize}. 
	
	 \begin{figure*} 
	  	\centering
	  	\includegraphics[width=16cm, trim={1cm 0cm 1cm 0cm}, clip]{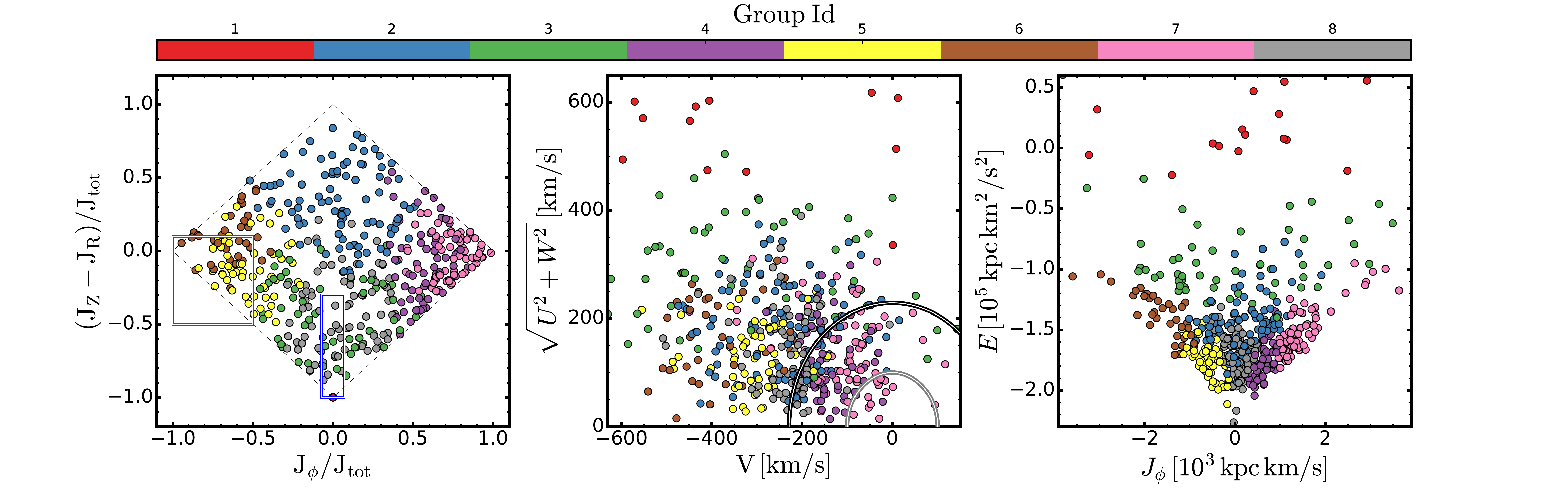}
	  	\includegraphics[width=15cm, trim={0cm 1cm 0cm 1cm}, clip]{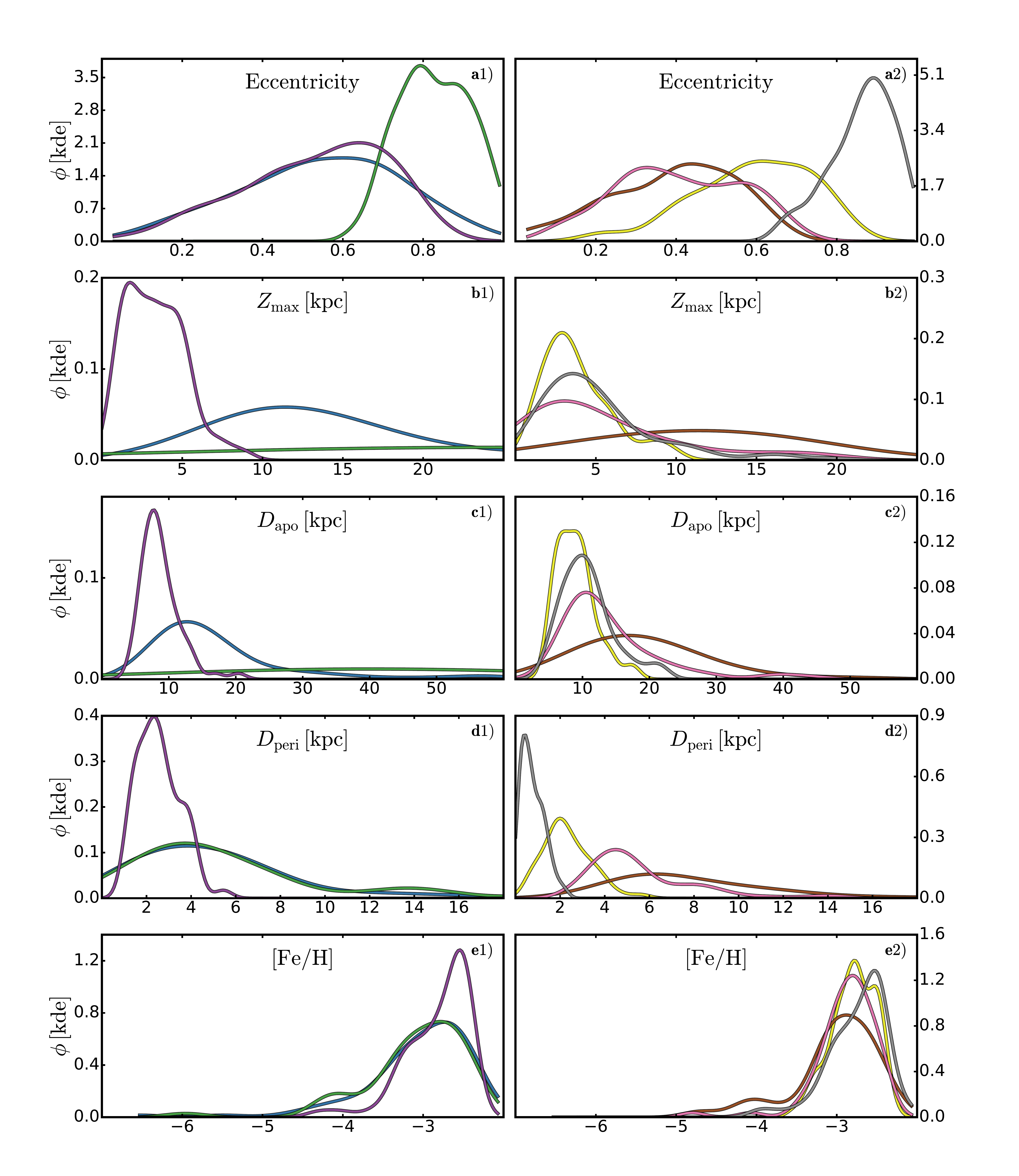}
	  	\caption{\textit{Top row.} Same as the top row Figure~\ref{fig:DiskAnalaysis} except stars are colour-coded by group.  The group colours are identified at the top of the panels. \textit{Panels a1)-e2).} Kernel density distributions of the computed orbital parameters for the groups identified by the clustering algorithm. In the panels with suffix 1 we represent the distributions for groups 2 to 4, while panels with suffix 2 show groups 5-8. The kernel density estimates of the orbital parameters for group 1 (G1) are not shown for scaling reasons.}
	  	\label{fig:clustering}
	 \end{figure*} 
	 
	 \begin{figure*} 
	  	\centering
	  	\includegraphics[width=15cm, trim={0cm 1cm 0cm 1cm}, clip]{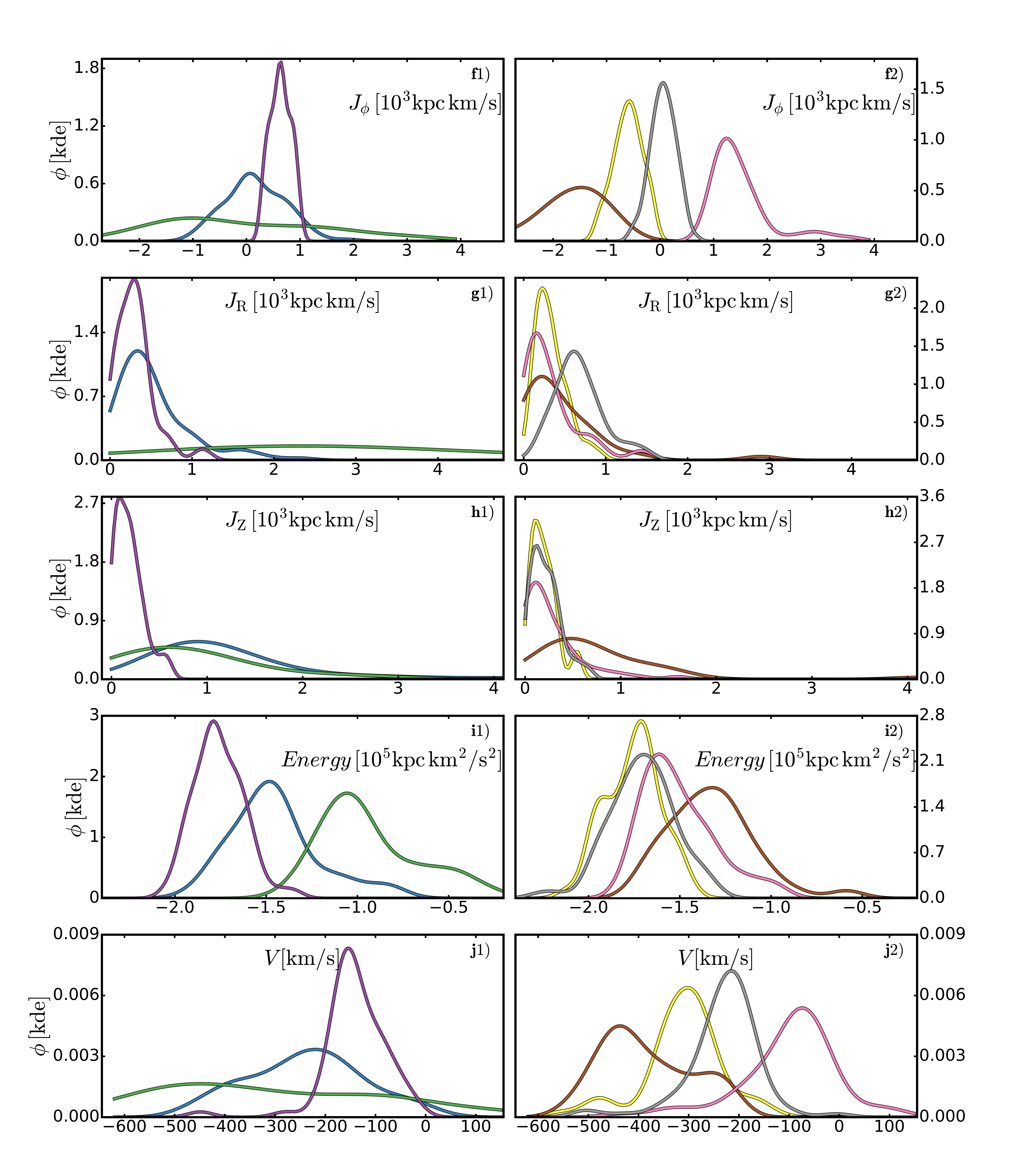}
	  	\caption{\textit{Panels f1)-j2).} Same as panels a1)-e2) of Figure~\ref{fig:clustering}.}
	  	\label{fig:clustering2}
	 \end{figure*} 

	Figure~\ref{fig:clustering} shows the results of the clustering analysis. In the top three panels we show the same top three panels as for Figure~\ref{fig:DiskAnalaysis} but now for the 8 identified clusters.  We note that group \# 1 (red markers with coordinates $(0, -1)$ in the action map) is the group of unbound stars, and therefore is not investigated further in the lower panels\footnote{For scaling reason some group \#1 stars are not shown in the top three panels of Figure~\ref{fig:clustering}.}.  Panels a1)--e2) in Figure \ref{fig:clustering} and panels f1) to j2) in Figure \ref{fig:clustering2} then show the kernel density distributions of different orbital parameters for each sub-group, with groups 2--4 in the left panels and groups 5--8 in the right panels.  
	A comparison of Figures~\ref{fig:clustering} and \ref{fig:clustering2} with Figure~\ref{fig:DiskAnalaysis} suggests the following.

	\begin{itemize}
		\item group \#~1 (G1, red markers, 28 stars) is composed of stars with energies consistent or greater than 0, visible in the top-right panel of Figure \ref{fig:clustering}. These stars have apparent apogalacticon distances larger than the Milky Way virial radius, i.e., 250\,kpc. They are characterized by large values of the radial action, $J_{\rm R}$, which translates in $J_{\rm \phi}/J_{\rm tot}\sim 0$ and $(J_{\rm Z}-J_{\rm R})/J_{\rm tot}\sim -1$. Overall, we find that 28 stars are grouped in G1. All of these stars are in the sub-sample discussed in Section~\ref{sub:escapers}, although two stars in the Section~\ref{sub:escapers} sub-sample, namely SMSS~J095211.09--185713.7 and SMSS~J002148.06--471132.1, are not classified as G1 stars, despite having apparent $D_{\rm apo}$  larger than 250\,kpc. These two stars have negative (bound) energies and are classified by the clustering algorithm in Group \#7 (pink points), which is composed of stars on loosely bound orbits. There is therefore essentially no discrepancy between the sub-sample discussed in Section~\ref{sub:escapers} and the high-energy G1 stars identified by the clustering algorithm.
		We do not show the kernel density distributions of these stars in the subsequent panels for scaling reasons. 
		\item group \#~2 (G2, blue markers, 93 stars) are a combination of prograde and retrograde halo stars. We note that they span quite a wide energy range, but we find difficult to draw further conclusions.  Presumably this group is made up of a mixture of in-situ and accreted halo stars.
	   	\item group \#~3 (G3, green markers, 54 stars) is a mixture of prograde and retrograde stars on loosely bound orbits, that venture far from the Galactic plane.
	   	\item group \#~4 (G4, purple markers, 61 stars) partially overlaps with low-$e$ stars in all three of the top-panels of Figure~\ref{fig:DiskAnalaysis}. Furthermore, their eccentricity distribution peaks at $e\sim0.4-0.6$, while most of them remain roughly confined within $5\,\rm kpc$ of the Galactic plane and 20\,kpc from the Galactic centre. A possible interpretation would be to consider these stars as thick-disk stars. This hypothesis is also supported by the Toomre diagram in the top center panel of of Figure \ref{fig:clustering}, where purple stars occupy a locus typical of thick-disk stars. 
	   	By the comparison with Figure~\ref{fig:DiskAnalaysis} we find a partial match of this group with the low-$e$ prograde population (orange points), identified as candidate very metal-weak thick disk stars.
		\item group \#~5 (G5, yellow markers, 53 stars) partially shares the location of the low-$e$ and     retrograde population (red dots in Figure~\ref{fig:DiskAnalaysis}) as well as partially overlapping with the locus of the Sequoia remnants identified in \citet{myeong2019}. The eccentricity distribution peaks at about $e\sim0.6$, and they are confined to the inner halo ($D_{\rm apo}\leq 10\, \rm kpc$). Panel b1) of Figure \ref{fig:clustering} shows that $\sim80\%$ of these stars are confined within $5\rm\,kpc$ from the Galactic plane. Most of them don't orbit further than 10 kpc from the Galactic centre.  Comparing then the top-right panels of Figure~\ref{fig:clustering} with \citet[][bottom-right panel of their Figure~2]{koppelman2019}, we see that G5 stars have a higher energy (in absolute values) than \textit{Gaia Sequoia} stars, while their energy and their angular momentum suggest a possible association with the Thamnos 1/2 groups \citep{koppelman2019}.   	
		\item group \#~6 (G6, brown markers, 37 stars) is composed of stars with very retrograde and mildly-eccentric orbits that venture far from the Galactic plane and from the Galactic centre, with $Z_{\rm max}$ and $D_{\rm apo}$ peaking at $\sim$15-20 kpc. These stars have energies that range from $\sim-1$ to $\sim-1.7\;[10^5\, \rm kpc\cdot km^2\,s^{-2}]$. This group is consistent with the identification of the \textit{Gaia Sequoia} remnants in \citet{koppelman2019}. 
		\item group \#~7 (G7, pink markers, 67 stars) partially overlaps with G4 both in the action map and in the Toomre diagram, while it is well-defined in the Energy vs. $J_{\rm \phi}$-plane. Panel a2) of Figure \ref{fig:clustering} shows that the eccentrity distribution of these stars is double peaked, with the first peak at $e\sim0.3$ and a second one at $e\sim 0.6$. Given the distribution of $D_{\rm apo}$ and $Z_{\rm max}$ it would seem that these stars are a mixture of halo and thick-disk stars, with lower binding energies than their G4 counterparts.
		As for G4, we note that there is a clear overlap between this group and the low-$e$ prograde stars shown in Figure~\ref{fig:DiskAnalaysis}.
		\item group \#~8 (G8, grey markers, 82 stars) share roughly the same location of the high-$e$ both retrograde and prograde population (azure and navy dots in Figure~\ref{fig:DiskAnalaysis}) in all top three panels of Figure \ref{fig:clustering}. Furthermore, the distribution of their orbital parameters nearly overlaps with those of the combined high-$e$ prograde and retrograde populations. Their energy and angular momentum agrees with the \textit{Gaia Sausage} definition in \citet{yuan2019}. We find particularly interesting the pericenter/apocenter distributions, whose analysis suggest that most of these stars move back and forth from the Galactic centre to the Galactic outskirts, always remaining within few $\rm kpc$ from the Galactic plane ($\sim$60\% these stars are indeed confined within $5\,\rm kpc$ from the plane). Given the observed orbital properties, and in particular the perigalacticon distances as low as $\sim$1 kpc, we speculate that such a remnant can be formed via a ``head-on'' accretion event, as in the {\it Sausage} progenitor \citep{myeong2019}.   

	\end{itemize}   

	It is clear that the two different analyses (discussed here and in Section~\ref{sec:discussion}) reach qualitatively the same conclusions. First, we find solid evidence for the existence of a very metal-weak component in the Galactic thick disk. 
	Further, from the analysis of the orbital actions and by means of the action map, we have identified possible members of the {\it Gaia Sequoia} and {\it Gaia Sausage} accretion events. The analysis also suggests that the low $|Z_{max}|$, high-$e$ population that is composed of stars with both prograde and retrograde orbits, may also be associated with the {\it Gaia Sausage} event. Both analyses also identify a consistent set of candidates that are likely not bound to the Galaxy.

\section{Example orbits}\label{app:orbits}	
	In the following we show four typical orbits of Halo, very metal-weak thick disk, \textit{Sequoia} and \textit{Sausage} stars within our sample.
	Each orbit is colour coded according to the integration time, while the white dot represents the current position in the Galactocentric cartesian reference frame. As discussed in Section~\ref{sub:orb}, each orbit has been integrated in a \texttt{McMillan2017} potential \citep{mcmillan2017} backward and forward in time for 2$\,\rm Gyr$. The actions have been computed with the St{\"a}ckel fudge method implemented in \texttt{GALPY}. %The observed parameters and the characteristics of the orbit are given in the leftmost panels for each star.
	 \begin{figure*}
	  	\centering
	  	\includegraphics[width=16cm, trim={7cm 0cm 0cm 1.5cm}, clip]{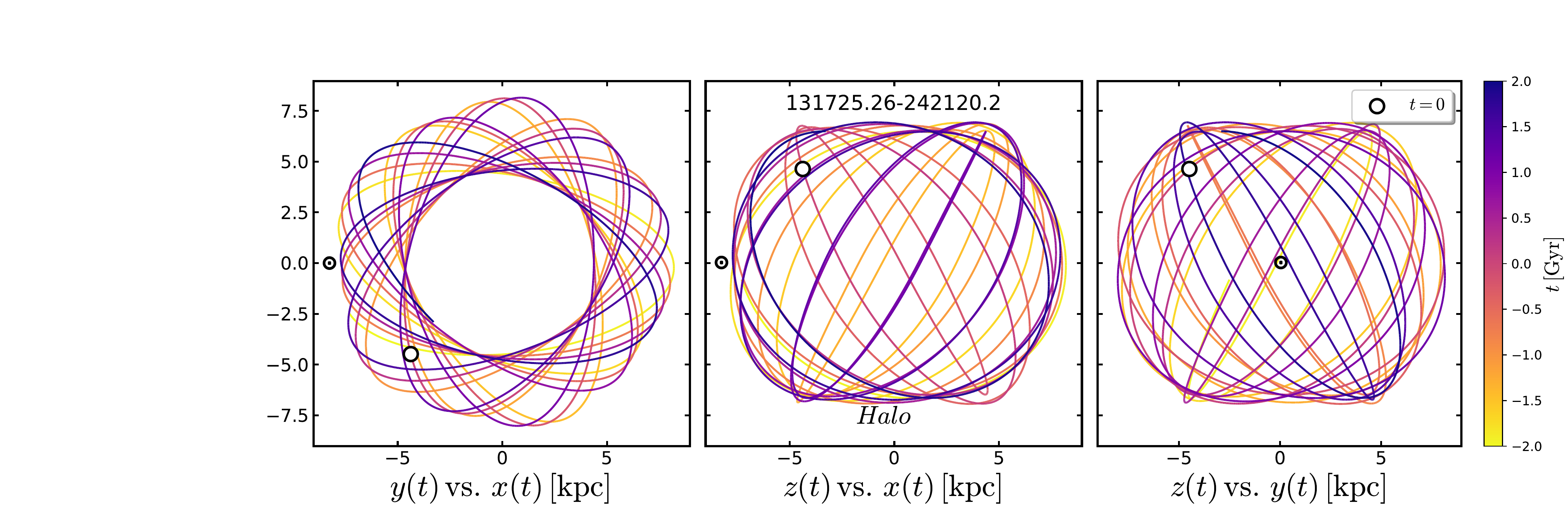}
	  	\includegraphics[width=16cm, trim={7cm 0cm 0cm 1.5cm}, clip]{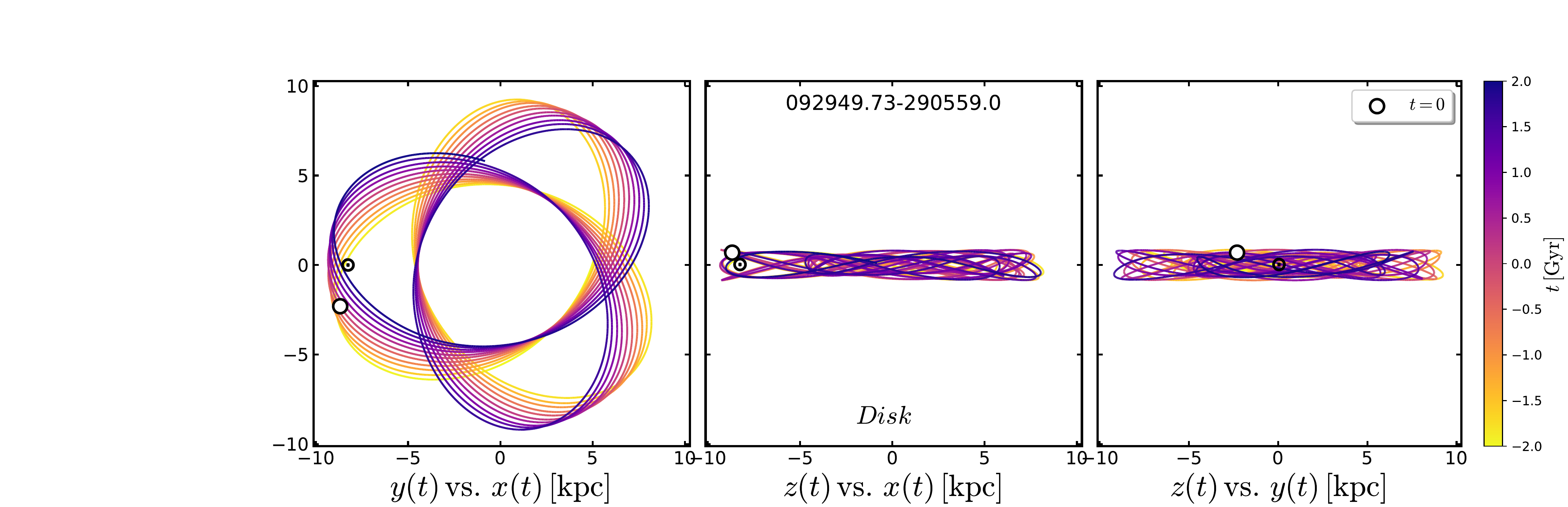}
	  	\includegraphics[width=16cm, trim={7cm 0cm 0cm 1.5cm}, clip]{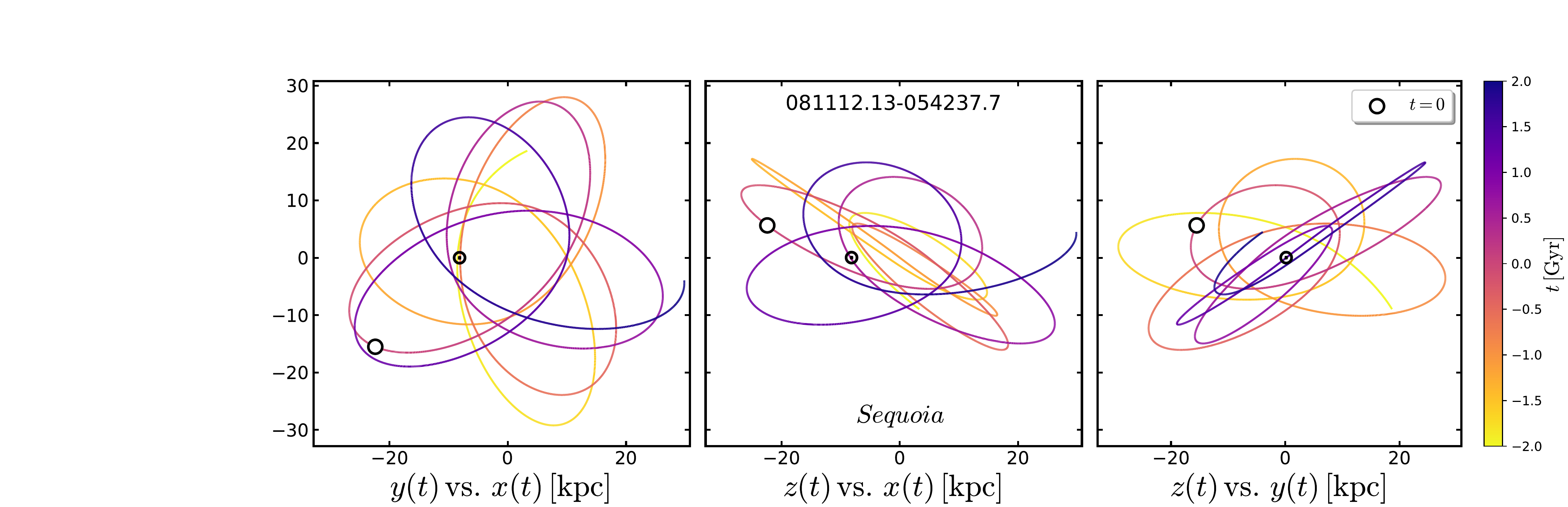}
	  	\includegraphics[width=16cm, trim={7cm 0cm 0cm 1.5cm}, clip]{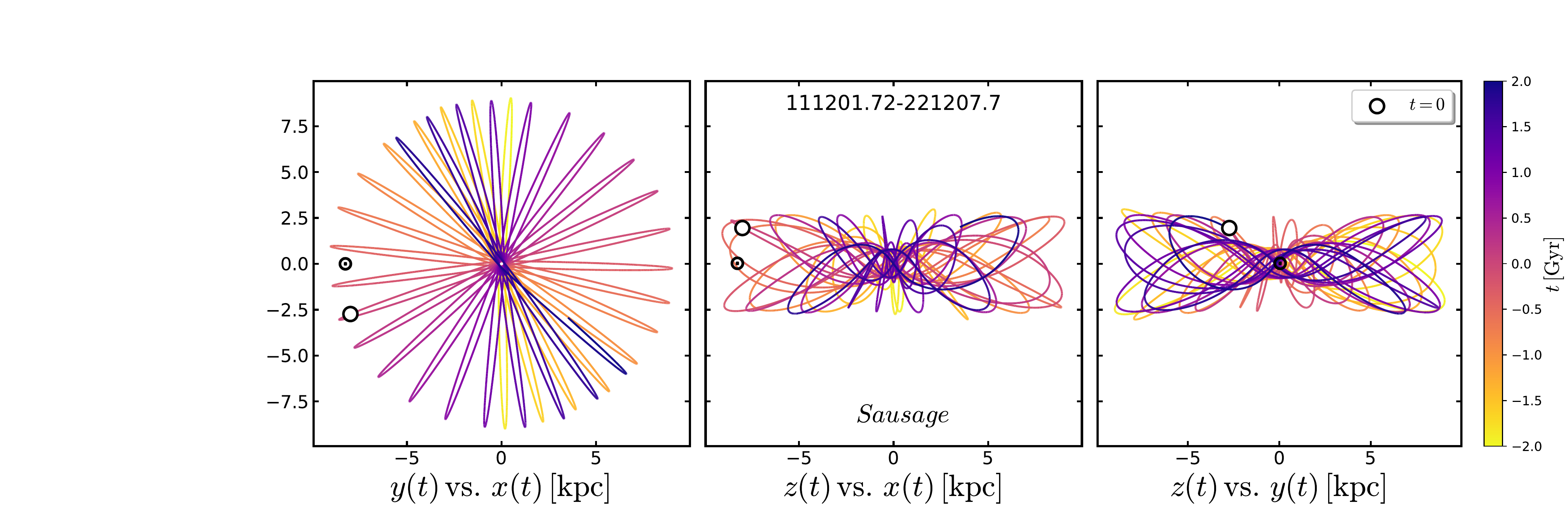}
	  	\caption{From top to bottom: typical orbit in the Galactocentric cartesian frame for examples of Halo, very metal-weak thick disk, \textit{Sausage} and \textit{Sequoia} stars, respectively. Each orbit is colour coded according to the integration time, and the white point indicates the current position of the star.  The position of the Sun is indicated by the circled dot.  Note that the orbit for the \textit{Sequoia} star shown in the third row is much larger than for the other three stars. }
	  	\label{fig:gSgEorbits}
	 \end{figure*}

%%%%%%%%%%%%%%%%%%%%%%%%%%%%%%%%%%%%%%%%%%%%%%%%%%

% Don't change these lines
\bsp	% typesetting comment
\label{lastpage}
\end{document}